\documentclass[conference]{IEEEtran}
\makeatletter
\def\ps@headings{%
\def\@oddhead{\mbox{}\scriptsize\rightmark \hfil \thepage}%
\def\@evenhead{\scriptsize\thepage \hfil \leftmark\mbox{}}%
\def\@oddfoot{}%
\def\@evenfoot{}}
\makeatother
\usepackage{subfigure}
\usepackage{times,amsmath,epsfig}
\usepackage{latexsym,amssymb}
\usepackage{graphicx,amssymb,amstext,amsmath}
\usepackage{setspace}
\usepackage{amsmath}
\usepackage{amsthm}
\usepackage{eucal}
\usepackage{amssymb}
\usepackage{mathrsfs}
\usepackage{color}
\usepackage{url}
\newtheorem{theorem}{Theorem}
\newtheorem{definition}{Definition}
\newtheorem{corollary}{Corollary}

\def\qed{\hfill{$\Box$} \\}
\def\eqdef{\triangleq}

\def\12{\frac{1}{2}}

\newfont{\bbb}{msbm10 scaled 500}

\newfont{\bb}{msbm10 scaled 1100}

\newtheorem{remark}{Remark}
 \newtheorem{example}{Example}

\IEEEoverridecommandlockouts

\begin{document}

\sloppy

\title{Physical Layer Security in Massive MIMO \thanks{This work was presented in part in the IEEE Computer and Network Security Conference (CNS), Florence, Italy, Septermber, 2015 }} 

\author{
  \IEEEauthorblockN{Y.~Ozan Basciftci}
  \and
  \IEEEauthorblockN{ C.~Emre Koksal}
  \and
  \IEEEauthorblockN{ Alexei Ashikhmin}

\thanks{Y. Ozan Basciftci and C.~Emre Koksal are with the Department of Electrical and Computer Engineering, The Ohio State University, Columbus, OH 43210, USA. Email: \{basciftci.1, koksal.2\}.osu.edu. Alexei Ashikhmin is with Bell Laboratories, Alacatel Lucent, Murray Hill, NJ, 07974, USA. Email: aea@research.bell-labs.com}
\thanks{This publication was made possible by NPRP grant 5-559-2-227 from
the Qatar National Research Fund (a member of Qatar Foundation) and ONR grant N00014-16-1-2253.}
  
}




\maketitle
\begin{abstract}
We consider a single-cell downlink massive MIMO communication in the presence of an adversary capable of jamming and eavesdropping simultaneously. We show that massive MIMO communication is naturally resilient to \emph{no training-phase jamming} attack in which the adversary jams \emph{only} the data communication and eavesdrops \emph{both} the data communication and the training. Specifically, we show that the secure degrees of freedom ($DoF$) attained in the presence of such an attack is identical to the maximum $DoF$ attained under no attack. Further, we evaluate the number of antennas that base station (BS) requires in order to establish information theoretic security without even a need for Wyner encoding. Next, we show that things are completely different once the adversary starts jamming the training phase. Specifically, we consider an attack, called \emph{training-phase jamming} in which the adversary jams and eavesdrops \emph{both} the training and the data communication. We show that under such an attack, the maximum secure $DoF$ is equal to zero. Furthermore, the maximum achievable rates of users vanish even in the asymptotic regime in the number of BS antennas.  To counter this attack, we develop a defense strategy in which we use a secret key to encrypt the pilot sequence assignments to hide them from the adversary, rather than encrypt the data. We show that, if the cardinality of the set of pilot signals are scaled appropriately, hiding the pilot signal assignments from the adversary enables the users to achieve secure $DoF$, identical to the maximum achievable $DoF$ under no attack. Finally, we discuss how computational cryptography is a legitimate candidate to hide the pilot signal assignments. Indeed, while information theoretic security is not achieved with cryptography, the computational power necessary for the adversary to achieve a non-zero mutual information leakage rate goes to infinity.
\end{abstract}
\section{Introduction}
\allowdisplaybreaks
Massive MIMO is one of the highlights of the envisioned 5G communication systems. In massive MIMO paradigm, the base station is equipped with a number of antennas, typically much larger than the number of users served. Combined with a TDD-based transmission, this solves many of the issues pertaining channel state information. In particular, the base station exploits law-of-large-numbers-like certainties as it serves each user over a combination of a large number of channels. 

While many issues behind the design of multicellular massive MIMO systems have been studied thoroughly, security of massive MIMO has not been actively addressed. Part of the reason for this may be the fact that, there is a vast literature on the security of MIMO systems in general, and a common perspective is that massive MIMO is merely an extension of MIMO as it pertains to security. However, we demonstrate that massive MIMO has unique vulnerabilities, and standard approaches to MIMO security do not address them directly. Instead, these approaches focus on issues that massive MIMO is naturally immune to. Furthermore, we argue that, common models used in MIMO security eliminate the need to think on various components of the system that are critical to understanding the vulnerabilities in security. In particular, in massive MIMO, merely making assumptions on available channel state information (CSI) is not sufficient, since the actual technique the system uses to obtain CSI may be the lead cause for some major security issues. For all these reasons, security of massive MIMO calls for a separate treatment of its own.

To that end, we consider the TDD-based single cell downlink massive MIMO system developed in~\cite{marzetta1} and later readdressed in~\cite{marzetta_conj}. The adversary is hybrid, capable of jamming and eavesdropping at the same time with its multiple antennas and we call our system {\em secure} if secrecy, measured in full equivocation is achieved at the adversary and arbitrarily low probability of decoding error is achieved at the legitimate receiver. We refer to these requirements as {\bf security constraints}. We first show how massive MIMO is naturally resilient to standard jamming and eavesdropping attacks, unless jamming is performed during the training phase when pilot signals are transmitted by the mobile users. We prove that, without pilot jamming, the achievable secure degrees of freedom\footnote{Our definition of degrees of freedom is different from the standard definition. Our definition specifies how the achievable rate scales with the log of the number of base station antennas, rather than the log of the transmission power as in the standard definition.}($DoF$) is identical to the maximum $DoF$ attained under no attack, even without the need to use a stochastic (e.g., Wyner) secrecy encoder in the massive MIMO limit. On the other hand, as we will show, the adversary can reduce the maximum secure $DoF$ and rate to zero by contaminating the pilot signal of the targeted user via another correlated pilot signal. To address this attack, we develop a defense strategy in which the base station (BS) keeps the assignment of pilot signals to the users hidden from the adversary and informs the assignments to the users reliably. Thus, in our approach, we use computational cryptography for encrypting the pilot assignments in the training phase. 
We also discuss how the consequences of encryption of pilot assignment is fundamentally different from the consequences of data encryption. In particular, we argue that, even if we use non-information theoretic methods (e.g., Diffie-Hellman) to encrypt the pilot assignments, the level of security we achieve can be as strong as information theoretic secrecy for all practical purposes. Note that, most of our results are {\bf not asymptotic} in the number of antennas and we specify the number of antennas necessary to achieve certain level of security.

The major ideas developed and demonstrated in this paper include:
\begin{itemize}
\item In information-theoretic secrecy literature, it is often the case that assumptions are made on the CSI available at the adversary. Typically, it is assumed that the adversary has access to the CSI for all channels in the system, with the motivation of making the achievable security robust with respect to the availability of CSI at the adversary. However, we show that, with massive MIMO, {\em it is not important {\bf if} the adversary has full CSI or not}. Indeed, we show that massive MIMO is naturally immune to attacks during data communication phase. Instead, we demonstrate that {\em the major question is {\bf how} the adversary obtains CSI.} In particular, we show that if the adversary is active during the training phase, it substantially degrades the security of data communication.
\item Security in computational cryptography is based on the assumptions on the computational power of the attackers. Once data is encrypted, it takes an unreasonable amount of time for a typical adversary to decrypt it without the key. Making such an assumption on the adversary poses a problem for security, since a sophisticated adversary can use various tools and techniques to cut down the time for cryptanalysis applied to recorded encrypted data. We eliminate this shortcoming by encrypting the {\bf pilot assignments} -not the transmitted data,- using keys that are shared via standard Diffie-Hellman. In our scheme, to make an impact, the adversary needs to decrypt the pilot assignment {\em before} the training phase starts. Note that, the training phase can start immediately after the assignments are made, leaving an arbitrarily low amount of time for the adversary to crack the assignment (i.e., pushing the computational power necessary to infinity). Without the knowledge of the pilot assignment, our scheme achieves perfect secrecy of information transmitted in the data communication phase, {\bf even without} the use of a secrecy encoder. Thus, it is useless for the adversary to record the received signal for future cryptanalysis, since it is indifferent from noise. 
\end{itemize}
Next, we summarize the technical contributions of our paper. Throughout the paper, we assume that the adversary is \emph{full-duplex}, i.e., it is capable of eavesdropping and  jamming the BS-to-user communication simultaneously. In the first part of the paper, we study an attack model in which the adversary eavesdrops the entire communication between the BS and users and jams only the downlink data communication (the adversary keeps silent during the training.). Under this attack:
\begin{itemize}
\item We show that the maximum secure $DoF$ is identical to the maximum $DoF$ achieved in the presence of no adversary.
\item We provide a novel encoding strategy, $\delta$-conjugate beamforming, that provides full security, without the need for  Wyner encoding~\cite{wyner1975}.
\item We evaluate the number of antennas that the BS requires in order to satisfy the security constraints.
\end{itemize} 
The proposed encoding, $\delta$-conjugate beamforming, utilizes the fact that the correlation between the estimated BS-to-user channel gains and the BS-to-adversary channel gains becomes \emph{zero} when the adversary does not jam during the training phase. We observe that in order to cause  a \emph{non-zero} correlation  between the estimated BS-to-user channel gains and the BS-to-adversary channel gains, the adversary has to jam the pilots of users. 

In the second part of the paper, we consider an attack model in which the adversary eavesdrops and jams the entire communication \emph{(including the training)} between the BS and the users. Under this attack:
\begin{itemize}
\item We show that, if the adversary jams the training such that there exists a \emph{non-zero} correlation between the BS-to-adversary channel gain and the estimated gain of the channel from the BS to a user, the adversary reduces the maximum secure $DoF$ to zero. Further, we show that, if the amount of the correlation is sufficiently large, the maximum achievable rate of the user also vanishes as the number of antennas at the BS grows.
\item We propose a counter strategy against the adversary. We show that, if the cardinality of the set of pilot signals scales with the number of antennas at the BS and the BS is able to keep the pilot signal assignments hidden from the adversary, attained secure $DoF$  is arbitrarily close to the maximum $DoF$ attained under no attack.
\end{itemize}

\emph{Related Work:} Massive MIMO concept was first proposed in~\cite{marzetta1,marzetta2}. Since then, there has been a flurry of studies focusing on different aspects of massive MIMO (see survey~\cite{tutorial}) such as channel estimation, energy efficiency, and pilot contamination. However, while MIMO security has been an active area of research \cite{contamination,energy_efficiency,blind_estimation}, issues specific to massive MIMO have not been considered. Among the very few, in~\cite{massive_secrecy}, the authors  consider downlink multi cell massive MIMO system in the presence of an adversary that only eavesdrops. In order to confuse the adversary, the BS transmits artificial noise from a set of its antennas. The authors conclude that, if the adversary has sufficiently large number of antennas, it is impossible to operate at a positive rate with artificial noise generation at the BS. In our earlier work~\cite{koksal_security}, which sets up the main results in this paper, we have focused on a fairly different model and addressed other questions. For instance, our attack model considers both jamming and eavesdropping, possibly simultaneously by the adversary.

\section{System Model and Problem Statement}\label{system_model}
\begin{figure}[t]
   \centering
   \includegraphics[width=0.50\textwidth]{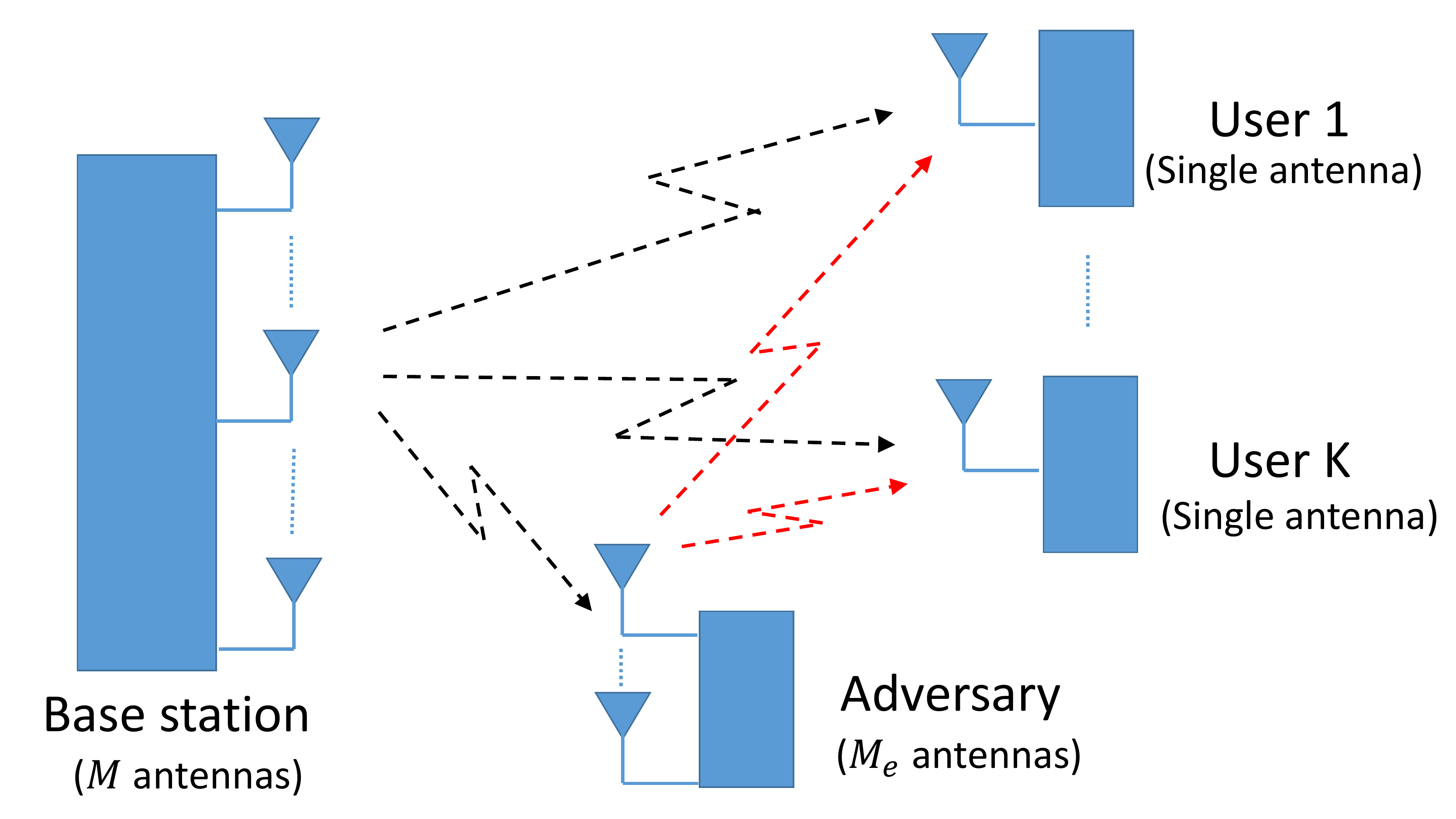}
   \caption{System Model}
   \label{fig:sysmdl}
 \end{figure}
We consider a multi user MIMO downlink communication system, depicted in Figure~\ref{fig:sysmdl}, including a base station (BS), $K$ single-antenna users, and an adversary. The BS equipped with $M$ antennas wishes to broadcast $K$ distinct messages $\left[W_1,\ldots,W_K\right]$ each of which is intended for a different user. The adversary is equipped with $M_e$ antennas. 
\subsection{Channel Model}
We assume all the channels in our system, illustrated in~Figure~\ref{fig:sysmdl}, are block fading. In the block fading channel model, time is divided into discrete blocks each of which contains $T$ channel uses. The channel gains remain constant within a block and the channel gains on different blocks are independent and identically distributed. Furthermore, we assume the channels are reciprocal; the instantaneous gain of the channel connecting the BS to a user is as same as the gain of the channel connecting to the same user to the BS.

We follow a TDD-based two-phase transmission scheme introduced in~\cite{marzetta_how} and is re-adressed in~\cite{marzetta_conj}. The signal transmission in a block is separated into two phases: training phase and data communication phase. On the first $T_r$ channel uses of every block,  each user sends a pilot signal to the BS. The BS estimates each BS-to-user channel from the observed pilot signals. On the last $T_d$ channel uses of each block ($T_d\eqdef T-T_r$), the BS transmits data to the users.

 The observed signals during a data communication phase at $k$-th user and at the adversary at a particular channel use of $i$-th block are as follows\footnote{Except for the channel gains, we avoid the block and channel use indices in~\eqref{l1} and~\eqref{l2} and the block indicies in \eqref{xx3} and \eqref{xx4} for the sake of simplicity.}: 
\begin{align}
Y_k& = H_k(i)X + H_{jam,k}(i)\color{black}V_{jam}+V_k\label{l1}\\
Z& =  H_e(i)X + V_e\label{l2},
\end{align}
where $Y_k$ is a received complex signal at $k$-th user, $Z$ is a received $M_e\times 1$ complex vector at the adversary, and $X$ denotes  $M\times 1$ complex vector of  transmitted data symbols. Signals $V_k$ and $V_e$ are  additive Gaussian noise components,  distributed as $\mathcal{CN}(0,1)$ and $\mathcal{CN}(\mathbf{0},I_{M_e})$, respectively. Signal $V_{jam}$ denotes $M_e\times 1$ complex vector of jamming signal. 
Further, $H_k(i)$ and $H_{jam,k}(i)$ denote a $1\times M$ complex gain vector of the channel connecting the base station to $k$-th user,  a $1\times M_e$ complex gain vector of the channel connecting the adversary to $k$-th user, respectively, at $i$-th block. Similarly, $H_e(i)$ is the $M_e\times M$ complex gain matrix of the MIMO channel connecting the base station to the adversary at $i$-th block. We assume that all channel gains $H_e(i), H_1(i),\dots H_K(i), H_{jam,1}(i),\dots, H_{jam,K}(i)$ are mutually independent for any $i\geq 1$. 

The users send pilots in the first $T_r$ channel uses of each block. The received signals at the BS and at the adversary in the training phase of $i$-th block are as follows:
\begin{align}
Y^{T_r} &=  \sum_{k=1}^K H_k^{\mathsf{T}}(i)\phi_k + H_e^{\mathsf{T}}(i)W_{jam}+W,\label{xx3}\\
Z^{T_r} &= \sum_{k=1}^K H_{jam,k}^{\mathsf{T}}(i)\phi_k + W_e,\label{xx4}
\end{align}
where $Y^{T_r}$ and $Z^{T_r}$ denote  $M\times T_r$ and $M_e\times T_r$ complex matrices of the received signals over $T_r$ channel uses at the BS and at the adversary, respectively. Signals $W$ and $W_e$ are $M\times T_r $ and $M\times T_r$ complex matrices denoting the additive Gaussian noise. Each element of $W$ and $W_e$ are i.i.d $\mathcal{CN}(0,1)$. Signal $V_{jam}$ denotes $M_e\times T_r $ complex matrix of jamming signal. Signal $\phi_k$ is $1\times T_r$ complex vector denoting the pilot signal associated with $k$-th user. The power of pilot signals  $\rho_r$, i.e., $\frac{1}{T_r}\text{tr}\left(\phi_k^{*}\phi_k\right)= \rho_r$ is  identical for all users $k\in\{1,\ldots, K\}$.

We assume that the users do not have the knowledge of the BS-to-user channel gains. Note that the BS, the users, and the adversary know pilot signal set $\left[\phi_1,\dots,\phi_K\right]$. The adversary is assumed to be aware of which pilot signal is assigned to which user. Utilizing the pilot signals, the BS estimates the BS-to-user channel gains. 
Define $\hat H_k(i)$  as $1\times M$ complex vector of estimated  BS-to-$k$-th user channel gain. 
Further, for any $B\geq 1$, define $H^B$, $\hat{H}^B$, $H^B_e$, and $H_{jam}^B$ as the gains of the BS-to-user channels, the estimated gains of the BS-to-user channels, the gains of the BS-to-adversary channel, and the gains of the adversary-to-user channels over $B$ blocks, respectively, i.e.,  $H^B\eqdef\left[H_1^B,\ldots, H_K^B\right]$, $\hat H^B\eqdef\left[\hat H_1^B,\ldots, \hat H_K^B\right]$ and $H_{jam}^B\eqdef \left[H_{jam,1}^B,\ldots,H_{jam,K}^B\right]$.

 For any $B\geq 1$, the joint probability density function of $\left(H^B, \hat H^B, H^B_e, H_{jam}^B \right)$ is 
\begin{align}
& p_{H^B,\hat H^B, H_e^B, H^B_{jam}}\left(h^B, \hat h^B, h_e^B,h_{jam}^B\right) =\nonumber\\
&\qquad\qquad \prod_{i=1}^B p_{H,\hat H, H_e, H_{jam}}\left(h(i), \hat h(i), h_e(i),h_{jam}(i)\right)\label{joint_dist}
\end{align}
where $H\eqdef \left[H_1,\dots, H_K\right]$, $\hat H\eqdef\left[H_1,\dots, \tilde H_K\right]$, and $ H_{jam}\eqdef\left[H_{jam,1},\dots, H_{jam,K}\right]$. For any $k\in\{1,\ldots, K\}$, $H_k$ and $H_{jam,k}$ are distributed as $\mathcal{CN}(\mathbf{0}, I_{M})$, $\mathcal{CN}(\mathbf{0}, I_{M_e})$, respectively, and each element of matrix $H_e$ is i.i.d $ \mathcal CN(0,1)$.  

The adversary has the perfect knowledge of the BS-to-user channel gains $H$ and the estimated BS-to-user channel gains $\hat H$. Define $H_{k_m}$ and $\hat H_{k_m}$ as the gain and the estimated gain of the channel connecting $m$-th BS antenna to $k$-th user. We assume that  for any $k\in\{1,\ldots, K\}$, $\{H_{k_m}\hat H_{k_m}\}_{m\geq1}$ forms an i.i.d process. We also assume that $\hat H_k$  are independent with $H_l$ and $\mathbb E\left[\hat H_k \hat H^{*}_l\right] = 0$ for $k\neq l$ and $k,l\in\{1,\ldots, K\}$. Note that we do not impose these assumptions for the BS-to-adversary channels.
\begin{remark}When MMSE estimator and mutually orthogonal pilot signals are employed at the BS for channel estimation, these assumptions are satisfied. \end{remark}

\subsection{Attack Model}\label{ad_model}
We consider a full duplex adversary that is capable of eavesdropping and jamming simultaneously. In the sequel, we consider two attack models that differ only in the adversary's jamming activity in the training phase.

In Section~\ref{clas_adv}, we consider an attack model in which the adversary jams \emph{only} during the data communication phase and eavesdrops both the training and the data communication phases. We call this attack model as \emph{no training-phase jamming}. In the no training-phase jamming, the adversary jams during the communication phase using a Gaussian jamming signal and keeps silent during the training phase. Specifically,  signal $W_{jam}$ in~\eqref{xx4} is identical to zero and jamming signal $V_{jam}$ in~\eqref{l1} is distributed as $\mathcal{CN}(\mathbf{0},\rho_{jam} I_{M_e})$, where $\rho_{jam}$ is the jamming power.

In Sections~\ref{sec:smart_jam} and \ref{sec:counter}, we consider an attack model in which the adversary jams and eavesdrops both the training and the data communication phases. We call this attack model as \emph{training-phase jamming}. The adversary strategy during the data communication phase in this attack model is the same as that described in the previous attack model (i.e., no training-phase jamming). Instead of jamming with random signals, the adversary jams during the training phase with structured signals. We provide a detailed description of the signals used for jamming the training phase in Section~\ref{sec:smart_jam} and~\ref{sec:counter}.
\subsection{Code Definition}
The BS aims to send message $w_k\in \mathcal W_k$, $k=1,\dots,K$, to $k$-th user over $B$ blocks with rate $R_k$, while keeping $w_k$ secret from the adversary. The BS and the users employ code $\left(2^{BTR_1},\dots, 2^{BTR_K}, BT_d\right)$ of length $BT_d$, that contains:\\
 \textbf{1)} $K$ message sets, $\mathcal W_k\eqdef \{1,\ldots,2^{BTR_k}\}, k=1,\dots,K$.\\
 \textbf{2)} $K$ injective encoding functions, $f_k$, $k=1,\dots, K$, where $f_k$ maps  $w_k\in\mathcal W_k$  to data signal sequence $s_k^{BT_d}\in \mathbb C^{BT_d}$ satisfying an average power constraint  such that 
\begin{equation}
\frac{1}{BT_d}\sum_{i=1}^B\sum_{j=T_r+1}^T \mathbb\vert s_k(i,j)\vert^2\leq \rho_k, \;\;k=1,\dots, K \label{power_constraint}
\end{equation} for all $w_k \in  \mathcal W_K$, where notation $(i, j )$ indicates the $j$-th channel use of $i$-th block, $\rho_k$ denotes the power constraint for $k$-th user, and $s_k(i,j)$ is the complex data signal of $k$-th user. Note that $\rho_f \eqdef \sum_{k=1}^K \rho_k$ is the cumulative average transmission power. Further, note that encoding functions, $f_k$, $k=1,\dots, K$ can be \emph{deterministic} or \emph{stochastic}. Codes using stochastic encoding functions referred to as stochastic codes and the ones using deterministic encoding functions are referred to as deterministic codes. \color{black}\\
\textbf{3)} Linear beamforming that maps data signals\footnote{Note that $s_k^{BT_d}\eqdef \{s_k(i,j)\}_{i=1:B, j=T_r+1:T}$ and notation $(\cdot)^{BT_d}$ applied to any variable has the same meaning.} $s_1^{BT_d}\times\dots\times s_K^{BT_d}$ to channel input\footnote{Note that  the channel input sequence satisfies the following average power constraint
\begin{equation}
\frac{1}{BT_d}\sum_{i=1}^B\sum_{j=T_r+1}^T \mathbb E\left[\vert\vert X(i,j)\vert\vert^2\right] \leq \rho_f \label{power2_constraint}
\end{equation}
for all $w_1\times \dots \times w_K \in \mathcal W_1\times \dots \times \mathcal W_K$, where the expectation is over estimated channel gains $\hat H$. The inequality~\eqref{power2_constraint} follows from the individual power constraint~\eqref{power_constraint} and from the fact that $\mathbb E\left[\hat H_k \hat H^*_l\right] =0$ for $k\neq l$} $X^{BT_d}$. Two beamforming strategies are used throughout the paper:
\begin{itemize}
\item \textbf{Conjugate beamforming}: When the BS employs conjugate beamforming, channel input at $j$-th channel use of $i$-th block can be written as
\begin{equation}
X(i,j) = \sum_{k=1}^K s_k(i,j) \frac{\hat H_k^*(i)}{\sqrt{M  \alpha_k}},\label{beamforming}
\end{equation}
for any $i\in\{1,\ldots, B\}$ and $j\in\{T_{\tau}+1,\ldots,T\}$, where $\alpha_k\eqdef \mathbb E\left[|\hat H_{k_m}|^2\right]$.
\item \textbf{$\delta$-conjugate beamforming}: We introduce a new beamforming strategy, called $\delta$-conjugate beamforming that is slightly modified version of conjugate beamforming. Let $\delta$ be a positive real number. When the BS employs $\delta$-conjugate beamforming, the channel input at $j$-th channel use of $i$-th block can be written as
\begin{equation}
X(i,j) = \sum_{k=1}^K s_k(i,j)\frac{\hat H_k^*(i)}{\sqrt{M^{1+\delta}\alpha_k}}. \label{re_state_beamforming}
\end{equation}
Note that, when $\delta=0$,  $\delta$-conjugate beamforming  becomes identical with conjugate beamforming in \eqref{beamforming}.
\end{itemize}
\textbf{4)} Decoding functions, $g_k$, $k=1,\ldots,K$, where $g_k$ maps $Y_k^{BT_d}$ to $\hat w_k\in\mathcal W_k$.

\subsection{Figures of Merit}\color{black}
We define the average error probability of  code $\left(2^{BTR_1},\dots, 2^{BTR_K}, BT_d\right)$  as
\begin{equation}
P_e \eqdef\mathbb P\left(\bigcup_{k=1}^K  g_k(Y^{BT_d}) \neq W_k\right),\nonumber
\end{equation}
where  $W_k$ is uniformly distributed on $\mathcal W_k$. 
We assume that the adversary targets a single user during  communication. 
The secrecy of the transmitted message for $k$-th user is measured by the equivocation rate at the adversary, which is equal to the entropy rate of transmitted message $w_k$ conditioned on the adversary's observations. 

\begin{definition}\label{def_rate}
A secure rate tuple $R_{1},\dots R_K$ is said to be achievable if, for any $\epsilon > 0$, there exists $B(\epsilon)>0$  and a sequence of  codes $\left(2^{BT R_1}, \dots, 2^{BT R_K}, BT_d\right)$ that satisfy the following:
\begin{align}
&P_e \leq \epsilon, \label{cond1}\\
&\frac{1}{BT} H\left(W_k| Z^{BT},H^B,\hat{H}^B,H^B_e \right) \geq R_k-\epsilon \label{l3}
\end{align}
for all $B\geq B(\epsilon)$ and $k\in\{1,\ldots, K\}$, where $Z^{BT}$ is the received signal sequence at the adversary over BT channel uses.
\end{definition}
We refer to the constraints in~\eqref{cond1} and~\eqref{l3}  as decodability and secrecy constraints, respectively. We also refer to both constraints as security constraints. We call the communication system information theoretically secure if both constraints are satisfied. Notice that the achievable rate tuple definition above is presented for a given $M$, i.e., $M$ remains constant for a sequence of codes $\left(2^{BTR_1},\dots, 2^{BTR_K}, BT_d\right)$, $B\geq B(\epsilon)$. 

In this paper, we  mainly focus on the massive MIMO limit. Specifically, we study on how achievable rate tuple $R_1,\dots,R_K$ behaves as $M$ goes to infinity. To that end, we use the following notion of degrees of freedom for each user.
\begin{definition}
A secure degrees of freedom tuple $d_1,\dots, d_K$ is said to be achievable, if there exists achievable rate tuple $R_1, \ldots R_K$ such that 
\begin{align}
d_k = \lim_{M\to\infty}\frac{R_k}{\log M},  \; k=1,\dots, K.  \label{cond3}
\end{align}
\end{definition}
In the literature, degrees of freedom is typically defined as the limit $\lim_{\rho_k\to\infty}\frac{R_k}{\log \rho_k}$. Since we aim to understand how $R_k$ changes with $M$ under constant $\rho_k$, the degree of freedom definition in~\eqref{cond3} is more relevant for our interest.

For a given achievable secure degrees of freedom tuple  $d_1,\dots, d_K$, we define the secure degrees of freedom of the downlink communication as the minimum value in the tuple, i.e., secure $DoF\eqdef \min_{k\in\{1,\ldots,K\}} d_k$. In the rest of the paper, when we use secure $DoF$, we mean secure degrees of freedom attained in the presence of an adversary, and when we use $DoF$, we mean degrees of freedom attained under no adversary.

In this paper, we characterize  the maximum secure $DoF$ in the presence of various security attacks described in Section~\ref{ad_model}.  Furthermore, we aim to develop defense strategies that achieve the maximum secure $DoF$ against the security attacks that would limit the maximum secure $DoF$ to zero, otherwise.

\section{Adversary not jamming The Training Phase }\label{clas_adv}
In this section, we show that downlink communication in a single-cell massive MIMO system is resilient to the adversary that jams only the data communication phase and eavesdrops both the communication and training phases. We show that the maximum secure $DoF$ attained under no training-phase jamming is identical to maximum $DoF$ attained under no adversary. Then, we show that we can establish information theoretic security without using stochastic encoding, e.g., Wyner encoding. Finally, we evaluate the number of antennas that BS needs to satisfy the security constraints without a need for Wyner encoding.
\subsection{Resilience of massive MIMO}
In this subsection, we evaluate the maximum secure $DoF$ of the downlink communication in the presence of \emph{no training-phase jamming}. 
Then, we show that the maximum secure $DoF$ attained in the presence of no training-phase jamming is as same as the maximum $DoF$ attained without an adversary. This result demonstrates the weakness of the no training-phase jamming in the massive MIMO limit. 

\begin{theorem}\textbf{(Maximum secure {DoF})}\label{thm:classical}
For given block length $T$ and data transmission phase length $T_d$, the maximum secure $DoF$ under no training-phase jamming  is given by $\frac{T_d}{T}$.\qed
\end{theorem}
The complete proof is available in Appendix~\ref{app:classical_attack}, where we first provide an upper bound on secure $DoF$ and then present a strategy to achieve the upper bound. 
Here, we provide a proof sketch. In order to find an upper bound on secure $DoF$, we consider a multiple output single output (MISO) communication system without an adversary, in which the BS communicates to a single user under  power constraint $\rho_f$. Further, we assume that the BS and the user have a perfect information of the channel gains. We show that the supremum of achievable rates leads to a secure $DoF$ of $\frac{T_d}{T}$. Hence, we conclude that  $\frac{T_d}{T}$ is an 
upper bound on secure $DoF$ attained in the multi user downlink communication model in Section~\ref{system_model}.

We now describe a strategy to attain the maximum secure $DoF$ in Theorem~\ref{thm:classical}. On the first $T_r$ channel uses of each block, the users send pilot signals that are mutually orthogonal. The BS uses minimum mean square estimator (MMSE) to estimate the BS-to-user channel gains.  The BS constructs $K$ codebooks, $\mathit{c_k}$, $k=1,\dots,K$, where  codebook $c_k$ contains $2^{BT\hat R_k}$ independently and identically generated codewords, $s_k^{BT_d}$ of length $BT_d$ and $\hat R_k > R_k$. The BS maps $k$-th user's message to a codeword with a stochastic mapping function $f_k$. Specifically, the BS maps message $w_k\in \left\{1,\ldots,2^{BTR_k}\right\}$ to randomized message $m_k\in \{1,\ldots,2^{BT\hat R_k}\}$ as in \cite{wyner1975} and then maps randomized message $m_k$ to one of the codewords in $\mathit{c_k}$, $k=1,\dots, K$.  Utilizing the conjugate beamforming in~\eqref{beamforming}, the BS maps $K$ codewords, $s^{BT_d}_k$, $k=1,\dots,K$ to channel input sequence $X^{BT_d}$. Each user employs typical set decoding~\cite{cover1991}. In order to show that secrecy constraint~\eqref{l3} for a particular user is satisfied, we give the adversary the other users' transmitted codewords.\qed

  In the next couple of remarks, we emphasize the robustness of the downlink communication system against no training-phase jamming.

\begin{remark}\textbf{(The weakness of the adversary not jamming the training phase)} In the proof of Theorem~\ref{thm:classical}, we show that $\frac{T_d}{T}$ is indeed an upper bound on the $DoF$ of a downlink communication without the presence of an adversary. Hence, with also showing that the secure $DoF$ of $\frac{T_d}{T}$ is attained in the presence of  the adversary, we conclude that 
no training-phase jamming attack does not degrade the performance of the communication in terms of $DoF$.  The reason that secure $DoF$ of $\frac{T_d}{T}$ is achieved is that the adversary keeps silent during the training phase; hence the estimated BS-to-user channel gains are independent with $H_e$. 

In the next section, we consider an adversary jamming the training phase. In the presence of such an adversary, the BS-to-user channel gains become correlated with $H_e$  and the maximum secure $DoF$ is reduced to zero.
\end{remark}

\begin{figure}
   \centering
   \includegraphics[width=0.5\textwidth]{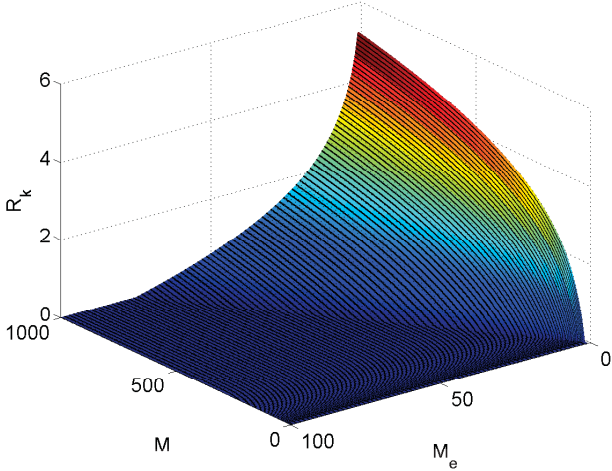}
   \caption{The variation of $R_k$ with $M$ and $M_e$}
   \label{fig:armrace1}
 \end{figure}

\begin{remark}\textbf{(Resource race between the BS and the adversary)}
In Appendix~\ref{app:classical_attack}, we show that the achievable rate tuple that leads to a secure $DoF$ of $\frac{T_d}{T}$ is $R_k =\frac{T_d}{T} \log\left(1+\frac{M\rho_ka}{\rho_f+\rho_{jam} +1}\right)-\frac{T_d}{T}\log\left(1+M_e\rho_k \right)$, $k=1,\dots,K$, where $a\eqdef \frac{\rho_rT_r}{\rho_rT_r+1}$. 

We next investigate how $R_k$ varies in  $M_e$ and $M$. Figure~\ref{fig:armrace1} illustrates this variation when $\rho_k=1$, $\rho_f=10$, $\frac{T_d}{T}=0.99$, $\rho_{jam}=1$, and $a=0.9$. As seen in Figure~\ref{fig:armrace1}, in the presence of the adversary not jamming the training phase, \textbf{the achievable secure rates are determined as a result of the arms race between the adversary and the BS.} Specifically, we can observe that if $M_e$ remains constant, achievable  rate $R_k$ grows unboundedly as $M$ is increasing. Moreover, for a fixed value of $M$, the achievable rates decrease as a function of $M_e$. In the next section, we consider an adversary jamming the training phase instead of keeping silent during the training phase. We will show that, armed with only a single antenna, the adversary is capable of limiting the maximum achievable rate for any user to zero as $M\to \infty$. \textbf{Hence, by jamming the training phase, the adversary converts the arms race between the BS and itself to the one between an user and itself.}
\end{remark}
\subsection{Establishing security without Wyner encoding}\label{subsection:nowyner}
In the achievability strategy given in the proof sketch of Theorem~\ref{thm:classical}, we use a stochastic encoding, a randomized mapping of each message to a codeword with stochastic functions, at the BS.  In fact, stochastic encoding, e.g., Wyner encoding~\cite{wyner1975}, is a standard technique in the literature for establishing information theoretic security against the eavesdropping attacks. 

In this section, we show that  the BS utilizing deterministic encoding, a nonrandom mapping of each message to a codeword with deterministic functions, instead of stochastic encoding is capable of satisfying the security constraints if it is equipped with sufficiently large number antennas. In order to satisfy the security constraints without using stochastic encoding, the BS employs novel beamforming strategy introduced in~\eqref{re_state_beamforming}.

The following theorem shows that, when code $\left(2^{BTR_1},\dots, 2^{BTR_K}, BT_d\right)$ of length $BT_d$ utilizes $\delta$-conjugate beamforming instead of conjugate beamforming in~\eqref{beamforming}, the code satisfies the secrecy constraint in~\eqref{l3}  for any $k\in\{1,\ldots,K\}$ and for any $\epsilon>0$ without a need for stochastic encoding.\color{black}
\begin{theorem}\textbf{(Establishing secrecy with no stochastic encoding)}\label{no_wyner}  Let $\delta>0$. Under no training-phase jamming,  for any $\epsilon>0$,  if $M\geq S(\epsilon)$, then any deterministic code $\left(2^{BTR_1},\dots, 2^{BTR_K}, BT_d\right)$  employing $\delta$-conjugate beamforming  satisfies
\allowdisplaybreaks
\begin{align}
\frac{1}{BT} H\left(W_k| Z^{BT_d}, H^B,\hat{H}^B,H^B_e \right) \geq R_k-\epsilon\label{norandom}
\end{align}
for all $B\geq 1$ and for all $k\in\{1,\ldots, K\}$, where $$S(\epsilon)\eqdef \left(\frac{M_e\rho_{max}}{2^{\frac{T}{T_d}\epsilon}-1}\right)^{\frac{1}{\delta}}$$ and $\rho_{max}\eqdef \max_{k\in \{1,\ldots, K\}}\rho_k$.\qed
\end{theorem}
We can consider $S(\epsilon)$ in Theorem~\ref{no_wyner} as the number of the  antennas the BS needs in order to make the conditional entropy $\epsilon$-close to $R_k$ for all $k\in\{1,\ldots, K\}$. Hence the BS equipped with at least $S(\epsilon)$ antennas  can satisfy~\eqref{norandom} by harnessing any code $\left(2^{BTR_1},\dots, 2^{BTR_K}, BT_d\right)$ that employs deterministic encoding functions and $\delta$-conjugate beamforming.

The proof is available in Appendix~\ref{no_wyner_proof}.  The BS constructs $K$ codebooks, $\mathit{c_k}$, $k=1,\dots,K$, where  codebook $c_k$ contains $2^{BTR_k}$ codewords, $s_k^{BT_d}$ of length $BT_d$. The BS maps message $w_k$ to  $s^{BT_d}_k$  codeword with a deterministic function, $f_k$, $k=1,\dots, K$. 
 Utilizing the $\delta$-conjugate beamforming in~\eqref{re_state_beamforming}, the BS maps $K$ codewords, $s^{BT_d}_k$, $k=1,\dots,K$ to channel input sequence $X^{BT_d}$.

\begin{figure}
   \centering
   \includegraphics[width=0.50\textwidth]{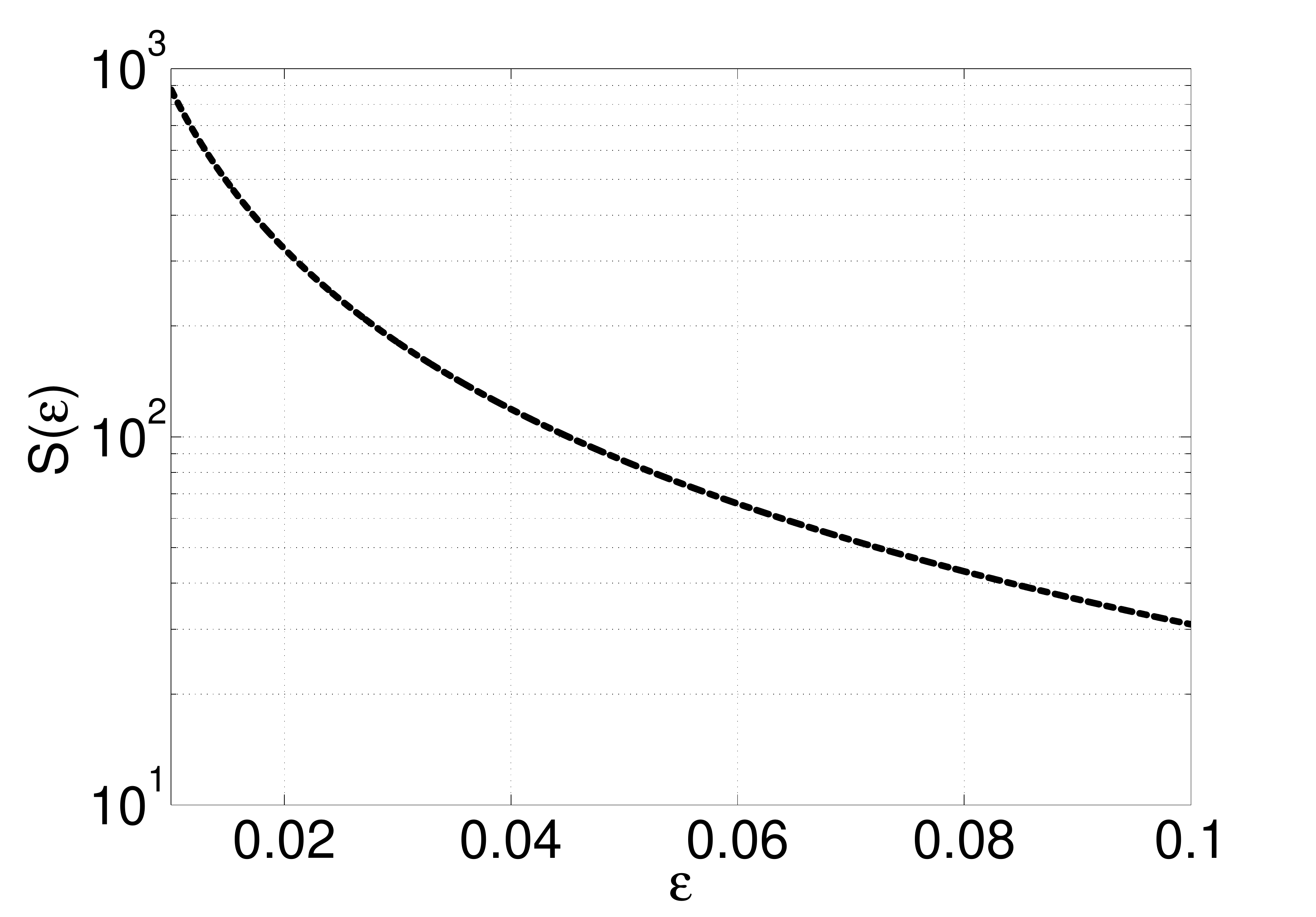}

   \caption{The variation of $S(\epsilon)$ with $\epsilon$ when  $ \rho_k =1$, $\delta=0.7$, $T/T_d=5/4$, and $M_e=1$. As long as $M\geq S(\epsilon)$, $\frac{1}{BT} H\left(W_k| Z^{BT_d}, H^B,\hat{H}^B,H^B_e \right)$ remains $\epsilon$-neighborhood of $R_k$ for any $k\in\{1,\ldots,K\}$.}
   \label{fig:massivechange}
 \end{figure}

In Figure~\ref{fig:massivechange}, we illustrate the variation of  $S(\epsilon)$ with $\epsilon$ when $ \rho_k =1$, $\delta=0.7$, $T/T_d=5/4$, and $M_e=1$. As seen in Figure~\ref{fig:massivechange}, $100$ antennas at the BS are sufficient to make the equivocation rate above $R_k-0.05$ for any choice of $R_k$, $k=1,\ldots,K$.

Theorem~\ref{no_wyner} evaluates the number of antennas needed in order to satisfy \emph{only} the secrecy constraint. The following corollary takes both the secrecy and decodability constraints into account.

\begin{figure}
   \centering
   \includegraphics[width=0.50\textwidth]{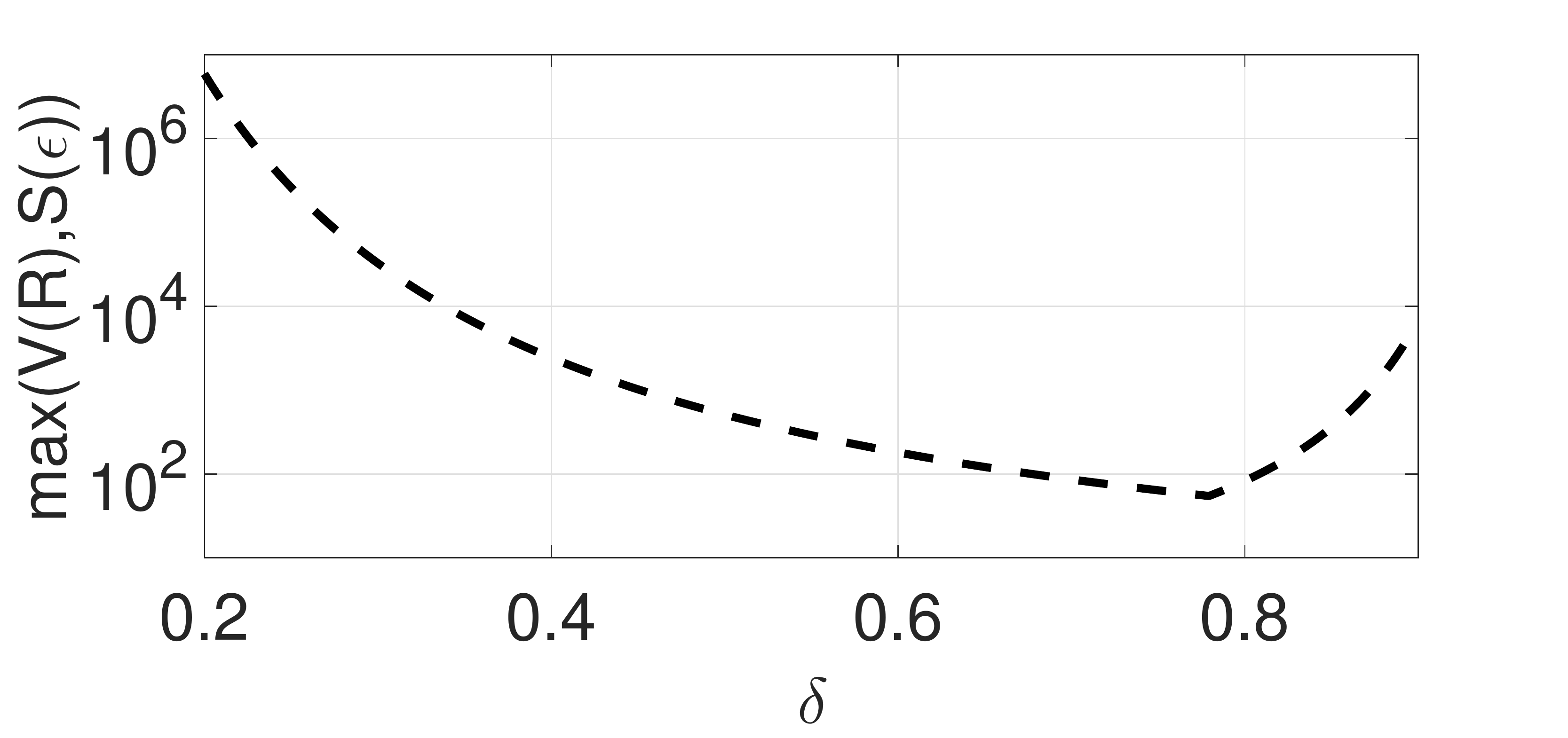}
   \caption{The variation of $\max\left(V(R),S(\epsilon)\right)$ with $\delta$ when $\epsilon=0.05$, $T/T_d=5/4$, $M_e=1$, $\rho_k =1$, and $R_k=0.2$ for any $k\in\{1,\ldots,K\}$. As long as $M\geq \max (S(\epsilon),V(R))$, constraints in~\eqref{cond1} and \eqref{l3} are satisfied for a given $\epsilon$ and $R$ \bf{without a need for stochastic encoding.}}
   \label{fig:max_S_eps_V_R}
 \end{figure}

\begin{corollary}\textbf{(Any rate tuple is achievable with no need to stochastic encoding)}\label{revisit}
 Let $0<\delta<1$.
In the presence of  no training-phase jamming,  for any $\epsilon>0$ and any rate tuple  $R\eqdef\left[R_1,\dots, R_K\right]$, if $M\geq \max\left(V(R),S(\epsilon)\right)$, there exists  $B(\epsilon)>0$  and sequence of  codes $\left(2^{BT R_1}, \dots, 2^{BT R_K}, BT_d\right), B\geq B(\epsilon)$ that satisfy the constraints in~\eqref{cond1} and \eqref{l3} \textbf{without the use of stochastic encoding}, where
$$
V(R) \eqdef \max_{k\in \{1,\ldots,K\}}\left(\left(2^{R_k\frac{T}{T_d}}-1\right)\times \frac{\rho_f+\rho_{jam}+1}{a\rho_k}\right)^{\frac{1}{1-\delta}}.
$$ \qed%
\end{corollary}
The proof of Corollary~\ref{revisit} can be found in Appendix~\ref{proof:revisit}. The sequence of codes in Corollary \ref{revisit} utilizes $\delta$-conjugate beamforming. Figure~\ref{fig:max_S_eps_V_R} illustrates the variation of $\max \left(V(R),S(\epsilon)\right)$ with $\delta$ when $\epsilon=0.05$, $T/T_d=5/4$, $M_e=1$, $\rho_k =1$, and $R_k=0.2$ for any $k\in\{1,\ldots, K\}$.  Note that, for these parameters, $\max   \left(V(R),S(\epsilon)\right)$ is minimized and identical to $60$ when $\delta=0.78$. When the BS utilizes $\delta$-conjugate beamforming with $\delta= 0.78$, the BS requires at least $60$ antennas in order to satisfy the constraints in~\eqref{cond1} and \eqref{l3}  without the need for a stochastic encoding  (e.g., Wyner encoding).
 
\begin{remark}\textbf{(Achieving secure DoF arbitrarily close the maximum DoF with no Wyner encoding)}
Theorem~\ref{no_wyner} and Corollary~\ref{revisit} show that it is possible to establish information theoretic security without using stochastic encoding. We next measure the amount of $DoF$ sacrificed as a result of not utilizing stochastic encoding. To that end, we evaluate how number of antennas at the BS  $\max(V(R),S(\epsilon))$ scales with $R_k$ for given $\epsilon>0$ and $\{R_l\}_{l\neq k}$. Specifically, we calculate $\lim_{R_k\to\infty}\frac{R_k}{\log\max(V(R),S(\epsilon))}$ as
\begin{align}
&\lim_{R_k\to\infty}\frac{R_k}{\log\max(V(R),S(\epsilon))}= \lim_{R_k\to\infty}\frac{R_k}{\log V(R)}\label{modified_dof1}\\
&=\lim_{R_k\to\infty}\frac{R_k}{\log\left(\left(2^{R_k\frac{T}{T_d}}-1\right)\times \frac{\rho_f+\rho_{jam}+1}{a\rho_k}\right)^{\frac{1}{1-\delta}}}\label{modified_dof2}\\
&=\lim_{R_k\to\infty}\frac{(1-\delta)R_k}{\log\left(2^{R_k\frac{T}{T_d}}-1\right)}\nonumber\\
&=(1-\delta)\frac{T_d}{T}\label{modified_dof3}
\end{align}
for all $k\in\{1,\ldots, K\}$, for any $\epsilon>0$ and for any $\{R_l\}_{l\neq k}$. The equalities in~\eqref{modified_dof1} and \eqref{modified_dof2} in the above derivation follow from the fact that $\max(V(R),S(\epsilon))$ and $V(R)$ are increasing functions of $R_k$. We observe from~\eqref{modified_dof3} that by choosing $\delta$  close to $0$, we can make the difference between \eqref{modified_dof3} and the maximum secure $DoF$ provided in Theorem~\ref{thm:classical} arbitrarily small.
\end{remark}
 \color{black}
\section{adversary jamming the training phase }\label{sec:smart_jam}
In the previous section, we show that the adversary not jamming during the training phase does not degrade the performance of the multi user communication when the BS has sufficiently large number of antennas. In this section, we aim to find attack model that do degrade the performance. Specifically, we focus on finding an attack strategy capable of limiting secure $DoF$ to an arbitrarily small value.  Next theorem sheds light on finding such an attack strategy. 
\begin{theorem}\label{zerodof}\textbf{(A non-zero correlation between the estimated user channel  and the adversary channel gains limits the maximum secure {DoF} to zero) }
Assume that there exists user $k$ such that
\begin{itemize}
\item $\left\{\hat H_{k_m}H_{e_m}\right\}_{m\geq1}$ is\footnote{$H_{e_m}$ is the gain of the connecting $m$-th antenna at the BS to the adversary.} an i.i.d random process. 
\item For any $B\geq 1$, there exists a random vector $\tilde H^B_k$ that satisfies the following: 1) the joint probability distribution of  $H^B_e, \hat H^B$ is identical with that of $H_k^B, \tilde H^B$, where $\tilde H^B\eqdef \hat H^B_1, \ldots, \tilde H^B_k,\ldots, \hat  H^B_K$ and 2) the joint probability distribution of $H(i), \tilde H(i)$ is identical for any $i\in [1:B]$.
\end{itemize} 
Then, the maximum secure $DoF$ is zero if  $\mathbb E\left[ H_{e_m}^{*}\hat H_{k_m}\right]\neq0$. \qed
\end{theorem}
Note that random vector $\tilde H^B$ is created by replacing $\hat H^B_k$ in $\hat H^B$ with $\tilde H_k^B$. The proof of Theorem~\ref{zerodof} can be found in Appendix~\ref{proof:zerodof}. In the example given at the end of this section, we show that the assumptions listed in Theorem~\ref{zerodof}, that are related to the random variables hold when MMSE and mutually orthogonal pilot signals are used as a channel estimation strategy. Note that such an estimation strategy is quite popular in the multi-user communication~\cite{marzetta2}. 

We next give a proof sketch.  Assume that conjugate beamforming is used at the BS and $M_e =1$.   Note that Theorem \ref{zerodof} is also valid when the BS uses $\delta$-conjugate beamforming and $M_e>1$.  \
We can convert the communication set-up explained in Section~\ref{system_model} to an identical set-up containing a BS equipped with $K$ antennas, where the channel input signal at $l$-th antenna in the new set-up represents the data signal for $l$-th user $S_l$, $l=1,\dots, K$.
Since conjuagte beamforming is used, the gain of the  channel connecting $l$-th antenna to $i$-th user in the new set-up is $\frac{H_i\hat H_l^*}{\sqrt{M\alpha_l}}$ and the gain of the  channel connecting $l$-th antenna to the adversary in the new set-up is $\frac{H_e\hat H_l^*}{\sqrt{M\alpha_l}}$, $i,l=1,\dots,K$. Following the assumptions in Theorem~\ref{zerodof}, we show that the gain of the  channel connecting the BS to the adversary can be  replaced with  $\frac{H_k \hat H^*_1}{\sqrt{M\alpha_1}},\ldots, \frac{H_k \tilde H^*_k}{\sqrt{M\alpha_k}},\ldots, \frac{H_k \hat H^*_K}{\sqrt{M\alpha_K}}$.   In Appendix~\ref{proof:zerodof}, we bound $R_k$ as follows
\begin{align}
&R_k\leq  \mathbb E\left[\left[\max_{\Sigma \in \mathcal S} \left(\log\left(1+A_k \Sigma A^{*}_k\right)\right.\right.\right.\nonumber\\
&\left.\left.\left.\qquad\qquad\qquad\qquad-\log\left(1+A_e\Sigma A^{*}_e\right)\right)\right]^{+}\right],\label{max_prob}
\end{align}
where $A_k\eqdef \left[\frac{H_k \hat H^*_1}{\sqrt{M\alpha_1}},\dots,\frac{H_k \hat H^*_k}{\sqrt{M\alpha_k}},\ldots, \frac{H_k \hat H^*_K}{M\alpha_K}\right]$ is $1\times K$ complex gain vector of channels connecting the BS to $k$-th user, and $A_e\eqdef \left[\frac{H_k \hat H^*_1}{\sqrt{M\alpha_1}},\ldots, \frac{H_k \tilde H^*_k}{\sqrt{M\alpha_k}},\ldots, \frac{H_k \hat H^*_K}{\sqrt{M\alpha_K}}\right]$ is $1\times K$ complex gain vector of channels connecting the BS to the adversary.
Let $\Sigma$ be the covariance matrix of input signal $S= \left[S_1,\dots, S_K\right]$ and $\mathcal{S}$ be the feasible set for the maximization problem in~\eqref{max_prob}. Every matrix $\Sigma$ in set $\mathcal{S}$ is diagonal due to fact that $S_1,\dots, S_K$ are independent, and satisfy $\Sigma \preceq diag (\rho_1,\ldots,\rho_k)$ due to the power constraint in~\eqref{power_constraint}.  

We show that, if $\mathbb E\left[H_{k_m}^{*}H_{e_m}\right]\neq0$, then the right hand side (RHS) of \eqref{max_prob} over $\log M$ goes to zero as $M\to \infty$. Hence, the maximum secure $DoF$ becomes zero. \qed
\begin{remark}\label{need_jam}(\textbf{Adversary has to jam the training phase})
When the adversary does not jam the training phase, $\hat H_k$ and $H_e$ are independent and consequently $\mathbb E \left[\hat H_{k_m}H^*_{e_m}\right] = \mathbb E\left[\hat H_{k_m}\right] \mathbb E \left[H^{*}_{e_m}\right]= 0$ for all $k\in \{1,\ldots, K\}$. In order to have a  non-zero correlation between the gain of the channel connecting itself to the BS $H_e$ with $\hat H_k$ for any $k\in\{1,\ldots, K\}$, the adversary has to jam the training phase. Hence, the training-phase jamming is capable of limiting the maximum $DoF$ to zero.\qed
\end{remark}
 In addition to limiting the maximum secure $DoF$ to zero, the adversary can make the maximum achievable rate of $k$-th user  arbitrarily small as $M\to \infty$.
We next provide the conditions under which the maximum achievable rate of $k$-th user goes to a finite value as $M\to \infty$.

\begin{corollary}\label{zerorate}\textbf{(A user's maximum achievable rate is bounded as $M\to \infty$)} In addition to the assumptions given in Theorem~\ref{zerodof}, assume that there exits  a finite non negative $r$ such that $p_{K_{M}}(x)\leq r$ for all $M\geq 1$ and $x\in\mathcal K_M$, where $p_{K_{M}}$ is the probability density function of $K_M\eqdef \frac{1}{M^2}||H_e\hat H^*_k||^2$ and $\mathcal K_M$ is the sample space of $K_M$. Then, the achievable rate of $k$-th user is bounded as 
\begin{equation}
\lim_{M\to\infty} R_k\leq \left[\log\left(\frac{\left\vert\mathbb E\left[H_{k_m}\hat H_{k_m}^{*}\right]\right\vert^2}{\left\vert\mathbb E\left[H_{e_m}\hat H_{k_m}^{*}\right]\right\vert^2}\right)\right]^+.\nonumber
\end{equation}
\end{corollary}\qed
The proof of Corollary~\ref{zerorate} can be found in Appendix~\ref{proof:zerorate}. As seen in Corollary~\ref{zerorate}, if the amount of correlation between the BS-to-$k$-user channel gain and the estimated BS-to-$k$-user channel gain, $\left\vert\mathbb E\left[H_{k_m}\hat H_{k_m}^{*}\right]\right\vert$ is smaller than that between the BS-to-adversary channel gain and estimated BS-to-$k$-th user channel gain, $\left\vert\mathbb E\left[H_{e_m}\hat H_{k_m}^{*}\right]\right\vert$, the maximum achievable rate of $k$-th user vanishes as $M\to\infty.$

\begin{remark}\textbf{(Resource race between the adversary and the user)} 
We show that if there exists a non zero correlation between the BS-to-$k$-user channel gain and the BS-to-adversary channel gain, then the maximum secure $DoF$ is constrained to zero. Furthermore, we also  show that if the amount of this correlation is higher than the amount of the correlation between the BS-to-adversary channel gain and estimated BS-to-$k$-user  channel gain, the maximum achievable rate of $k$-th user goes to zero as $M\to \infty$.

\textbf{Hence, in the presence of the training-phase jamming, the achievable rates and the maximum secure DoF are determined as a result of the arms race between the adversary and users.}\qed
\end{remark}
\begin{example} \textbf{(Using MMSE and mutually orthogonal pilot signals for channel estimation)} We study an adversary that chooses to match $k$-th user's pilot signal on the training phase with one of its antennas when  MMSE and mutually orthogonal pilot signals are used for channel estimation.  We  show that the assumptions  given in Theorem 3 are valid under such a jamming attack and a channel estimation strategy. Then, we show that the maximum secure $DoF$ is zero. 

We consider  mutually orthogonal pilot signals  $\{\phi_l\}_{l\in[1:K]}$, i.e.,
\begin{equation}
\phi_k \times \phi_l^{*}=
\begin{cases} 
T_r \rho_r&\mbox{if } k = l \\
0& \mbox{if }k\neq l
\end{cases}\nonumber
\end{equation} 
for any $k,l \in \{1,\ldots, K\}$. The received signals at the BS in the training phase of $i$-th block is as follows:
\begin{align}
Y^{T_r} = \sqrt{\frac{\rho_{jam}}{\rho_r}}H_e^{\mathsf{T}}(i)\phi_k+ \sum_{l=1}^K H_l^{\mathsf{T}}(i)\phi_l + W,\nonumber
\end{align}
where $\rho_{jam}$ is the jamming power. Note that we assume that the adversary jams the data communication phase and the training phase with the same power, which is $\rho_{jam}$.  

In order to validate the assumptions listed in Theorem~\ref{zerodof}, we next present the estimated gain of the channel connecting the BS to $l$-th user at $i$-th block as
\begin{align}
\hat H_{l}(i)=
\begin{cases} 
aH_{l}(i)+bH_{e}(i)+cV_l&\mbox{if } l = k \\
dH_{l}(i) + eV_l & \mbox{if }l\neq k,
\end{cases}
\nonumber
\end{align}
where $V_l$ is distributed as $\mathcal{C}\mathcal{N}(0,I_{M})$ for any $l\in \{1,\dots,K\}$, $a\eqdef \frac{T_r\rho_r }{T_r\rho_r  +1+T_r\rho_{jam}}$, $b\eqdef \frac{T_r\sqrt{\rho_r \rho_{jam}}}{T_r\rho_r  +1+T_r\rho_{jam}}$, $c \eqdef\frac{\sqrt{T_r\rho_r }}{T_r\rho_r +1+T_r\rho_{jam}}$, $d \eqdef \frac{T_r\rho_r }{T_r\rho_r  +1}$
 and $e \eqdef  \frac{\sqrt{T_r\rho_r }}{T_r\rho_r  +1}$.

Define $\tilde H^B_k$ stated in Theorem~\ref{zerodof} as $\tilde H_k(i)\eqdef  bH_k(i) + a H_e(i)+c V_k$, $i=1,\dots, B$. Further, define $\hat H_l\eqdef aH_{l}+bH_{e}+cV_l $ if $k=l$, and otherwise, $\hat H_l\eqdef dH_{l}+eV_l$. Note that $\tilde H_k(i),H(i), H_e(i),\hat H(i)$ is an i.i.d process due to~\eqref{joint_dist} and the associated joint distribution  is identical with that of $\tilde H_k,H, H_e,\hat H$, where $\tilde H_k\eqdef aH_{k}+bH_{e}+cV_k $. Hence, we conclude that the joint probability  distribution of $H(i), \tilde H(i)$ is identical for any $i\in \{1,\ldots,B\}$. 

We next show that  the  probability distribution of $H^B_e, \hat H^B$ is identical with that of $H_k^B, \tilde H^B$. Note that  both $(H_e, \hat H_k)$  and $(H_k,\tilde H_k)$ are independent with $\{\hat H_l\}_{l\neq k}$. Hence, noting that $H_e$ and $H_k$ have same probability distributions, it is sufficient to show that $\tilde H_k|H_k=h_k$ has the same probability distribution with $\hat H_k|H_e=h_k$ for any $h_k\in\mathbb R^M$:
\begin{align}
&\mathbb P\left(\tilde H_k\leq x| H_k=h_k\right)\nonumber\\
&=\mathbb P\left(b h_k +a H_e+cV_k\leq x| H_k=h_k\right)\nonumber\\
&=\mathbb P\left(b h_k +a H_e+cV_k\leq x\right)\label{zz3}\\
&=\mathbb P\left(b h_k +a H_k+cV_k\leq x\right)\label{zz4}\\
&=\mathbb P\left(b H_e +a H_k+cV_k\leq x|H_e=h_k\right)\label{zz5}\\
&=\mathbb P\left(\hat H_k\leq x| H_e=h_k\right)\nonumber
\end{align}
for any $x\in \mathbb R^M$, where \eqref{zz3} and \eqref{zz5} follow from the fact that $H_e$, $H_k$, and $V_k$ are mutually independent and \eqref{zz4} follows from the fact that  $\left(H_e,V_k\right)$ and $\left(H_k,V_k\right)$ are identically distributed.

Finally, note that $\{H_{e_m}, \hat H_{k_m}\}_{m\geq 1}$ forms an i.i.d process due to the fact that $H_k, H_e, V_k$ are mutually independent random vectors and each is composed of $M$ i.i.d complex Gaussian random variables.  

Note that $\mathbb E\left[\hat H^*_{k_m}H_{e_m}\right]=b$. Since $\mathbb E\left[\hat H^*_{k_m}H_{e_m}\right]$ is non-zero, we conclude that the maximum secure $DoF$ is zero by Theorem~\ref{zerodof}. \qed
\end{example}

\section{ Secure communication under Training-Phase Jamming}\label{sec:counter}
In the previous section, we showed that massive MIMO systems are vulnerable to the training-phase jamming. In this section, we first provide a defense strategy against the training-phase jamming, that expands the cardinality of the set of pilot signals and keeps the pilot signal assignments hidden from the adversary. Then, we show that utilizing the defense strategy and $\delta$-conjugate beamforming, the BS can satisfy the security constraints without using Wyner encoding in the presence of the training-phase jamming. Finally, we discuss that relying only on the computational cryptography, we can secure the communication of pilot signal assignments; hence the entire massive MIMO communication.
\subsection{Counter strategy against training-phase jamming}\label{sub_section:counter}
We first describe our defense strategy against training-phase jamming attack. Then, in Theorem~\ref{encrypting},  we show that the ratio of the achieved rate to the logarithm of number of antennas can be brought arbitrarily close to maximum achievable secure $DoF$ of $\frac{T_d}{T}$ with the proposed defense strategy that will be explained next.

The BS constructs pilot signal set $\Phi $ containing $L$ mutually orthogonal pilot signals, i.e., $\Phi =\left\{\phi_1,\ldots, \phi_L\right\}$, where $L$ is larger than the number of users in the system, $L\geq K$. Thus, the number  the pilot signals is increased. At the beginning of each block, the BS draws $K$ pilot signals from set $\Phi$ uniformly at random and assigns each of them to a different user.  Let $\Phi_K(i) = \left[\phi_{1}(i),\dots,\phi_{K}(i)\right]$ be $K$ pilot signals that the BS picks at the beginning of $i$-th block, where $\phi_{k}(i)\in \Phi$  is the pilot signal assigned to $k$-th user on $i$-th block. 

Throughout sections~\ref{sub_section:counter} and \ref{sub_section_comput}, we assume that the BS communicates to the users the assignments of pilot signals reliably while keeping the assignments hidden from the adversary. In Section~\ref{subsection:share_securebits}, we discuss how this can be achieved. In particular, we consider \emph{computational cryptography} as a way to communicate the pilot signal assignments and discuss the notion of security achieved. 

We next describe the attack model in detail, under the lack of knowledge of the pilot signal assignments. Suppose that the adversary targets $k$-th user without loss of generality. The adversary eavesdrops the entire communication between user $k$ and the BS and simultaneously jams data communication phase with Gaussian noise as in no training-phase jamming attack model. Furthermore, the adversary, \emph{without knowing which pilot signal is assigned to which user},  picks $J\leq L$ pilot signals uniformly at random from set $\Phi$ at the beginning of a block and subsequently jams these pilot signals with an equal power during the training phase.  The adversary repeats this process independently at the beginning of each block.

Particularly, the adversary divides its jamming power and transmits an equally weighted combination of $J$ randomly selected pilot signals using all of its $M_e$ antennas with total transmission power $\frac{\rho_{jam}}{J}$. The signal received by the BS during the training phase under this attack model can be written as follows:
\begin{align}
Y^{T_r} = \sum_{l=1}^K H_l^{\mathsf{T}}\phi_l +\sum_{l\in\mathcal J}\sum_{n=1}^{M_e}\sqrt{\frac{\rho_{jam}}{M_eJ\rho_r}}H_{e_n}^{\mathsf{T}}\phi_l + W \label{l333},
\end{align}
where $Y^{T_r}$ denotes  $M\times T_r$ complex matrix of the received signals over $T_r$ channel uses at the BS,  $H_{e_n}$ is  $1\times M$ complex gain vector of the channel connecting  $n$-th antenna at the adversary to the BS, and $\mathcal J$ is the set of  pilot signals that are selected  and transmitted by the adversary at the corresponding block. Note that ${\mathcal J}$ is a random set that can possibly change in each block and $|{\cal J}|=J$.

Next theorem shows that when the cardinality, $L$ of pilot signal set is increased as a function of the number of BS antennas in a certain way, the ratio of attained secure rate to  $\log M$ for any user can be arbitrarily close to the maximum $DoF$ attained in the presence of no adversary. 

\begin{theorem}\label{encrypting}\textbf{(Achievable rate under training-phase jamming)}  For given block length $T$ and data transmission phase length $T_d$, the achievable secure rate, $R_k$ under  training-phase jamming satisfies 
\begin{align}
\frac{R_k}{\log M}\geq  \frac{T_d}{T} \min(1,\gamma)  -\epsilon \label{eq:lower_bound}
\end{align}
for any $k\in \{1,\dots,K\}$,  $J\in \{1,\dots T_r\}$,  $\epsilon>0$, and $\gamma>0$ if  $ \max(M^{\gamma},K)\leq T_r $ and $M\geq G(\epsilon)$, where  
\begin{align}
&G(\epsilon)\eqdef\left(\left(1+M_e\rho_{max}+\frac{M_e\rho_{max}\rho_{jam}}{\rho_r}\right)\right.\nonumber\\
&\;\;\qquad\qquad\left. \times(\rho_f+\rho_{jam} +1)\times\frac{\rho_r+\rho_{jam}+1}{\rho_{min} \rho_r}\right)^{\frac{T_d}{T\epsilon}},
\end{align}
$\rho_{max}\eqdef \max_{k\in \{1,\dots,K\}}\rho_k$, and $\rho_{min}\eqdef \min_{k\in \{1,\dots,K\}}\rho_k$.
\qed
\end{theorem}
Note that the lower bound to $\frac{R_k}{\log M}$ in~\eqref {eq:lower_bound} does not depend on how many pilot signals the adversary chooses to contaminate. 
The proof of Theorem~\ref{encrypting} can be found in Appendix~\ref{proof_of_encrypting}.
\begin{remark}\textbf{(Attained $\frac{R_k}{\log M}$ is arbitrarily close to maximum $DoF$ under no attack)} 
We can observe from the statement of Theorem~\ref{encrypting} that  when $\gamma = 1$, $\frac{R_k}{\log M}$ that is arbitrarily close to the maximum $DoF$ attained under no attack can be achieved.  In order to attain that amount of $\frac{R_k}{\log M}$, the length of the training phase $T_r$ is expanded so that $T_r \geq \max (K, G(\epsilon))$ for given $\epsilon>0$ and the size of pilot signal set is set to $T_r$ instead of $K$.  Hence, we sacrifice the some of secure throughput by increasing the training overhead. However, as illustrated in the next example, the typical values for the block lengths for mobile wireless communication systems is sufficiently large to keep the overhead ratio, $\frac{T_r}{T}$ reasonably low.
\end{remark}

\begin{example} In this example, we consider massive MIMO downlink transmission to users moving at a speed 10 m/s and the transmitted signal bandwidth is 10 MHz, centered at 1 GHz The associated coherence time corresponds $T$ to as $3\times 10^5$ channel uses.  We first evaluate the number of antennas required to keep $R_k$ in $\epsilon$ neighborhood of $\frac{T_d}{T}$ for a given training phase length $T_r$. To that end, we plot the variation of $G(\epsilon)$ with $\epsilon$ in Figure~\ref{fig:gamma}, when $\gamma =1$, $T_d= 2\times10^5$ channel uses, $p_{jam}=1$, $K = 5$, $p_f = 5$, $M_e =1$, $p_r=10$, $p_k =1$ for all $k\in\{1,\ldots, K\}$. For these set of parameters, 200 antennas are sufficient to keep $\frac{R_k}{\log M}$ larger than $\frac{T_d}{T}-0.3$, where $\frac{T_d}{T} = \frac{2}{3}$. 

Next we study the trade-off between $\epsilon$  and $\frac{T_d}{T}$ for given $M$, where $\epsilon$ is the deviation of achieved $\frac{R_k}{\log M}$ from $\frac{T_d}{T}$ as in \eqref{eq:lower_bound}. To that end, we plot the variation of $\epsilon$ and $\frac{T_d}{T}$ with $\frac{T_r}{T}$ for $M=200$ as we change $\frac{T_r}{T}$ from $\frac{200}{3\times 10^5}$ to $1$. The values of parameters $p_{jam}$, $K$, $M_e$, $\rho_k$, and $\rho_f$ are kept same as stated above and we set $\rho_r = \frac{T_d}{T_r} \rho_f$. As seen in Figure~\ref{fig:tradeoff}, $\epsilon$ vanishes as $T_r$ goes to $T$ and hence $\frac{R_k}{\log M}$  also gets closer to  $\frac{T_d}{T}$. However, as $T_r$ increases, the training overhead increases and hence maximum $DoF$ $\frac{T_d}{T}$ decreases. 

\begin{figure}[t]
   \centering
   \includegraphics[width=0.50\textwidth]{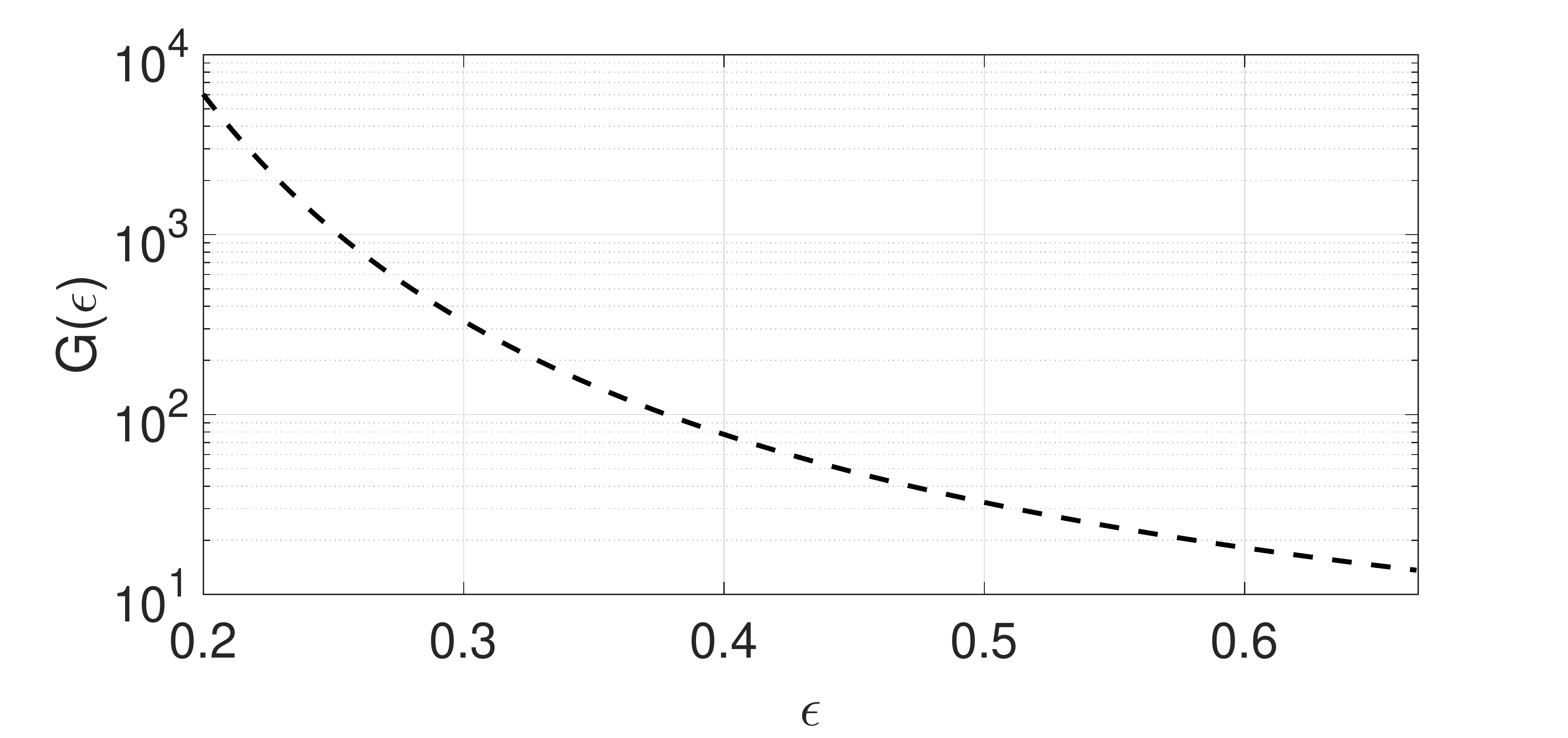}
   \caption{The change of $G(\epsilon)$ with $\epsilon$}
\label{fig:gamma}
 \end{figure}
\begin{figure}[t]
   \centering
   \includegraphics[width=0.50\textwidth]{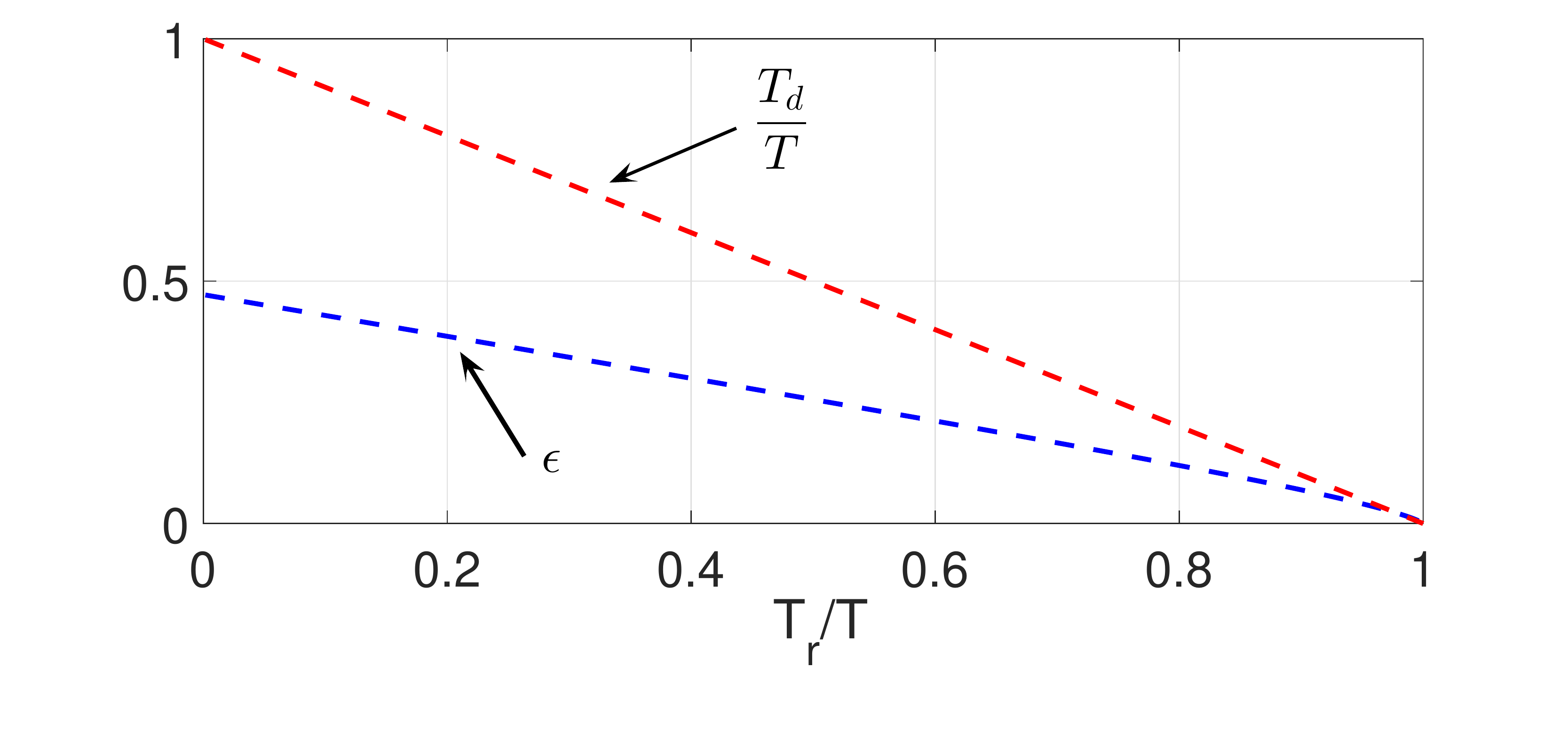}
   \caption{The change of $\epsilon$ in~\eqref{eq:lower_bound} and $\frac{T_d}{T}$ with $\frac{T_r}{T}$}
\label{fig:tradeoff}
 \end{figure}
\end{example}

\begin{remark} \textbf{(Resource race between the adversary and the BS)} By keeping the pilot assignments hidden from the adversary and using a pilot signal set that scales with $M$, \textbf{the BS converts the arms race between the adversary and the target user (which was the case with known pilot assignments), back  to the one between the adversary and itself.} Indeed, the power of the adversary needs to scale with $L$ for it to make an impact.
\end{remark}
\subsection{Establishing security without Wyner encoding}\label{sub_section_comput}
In this subsection, we show that the BS, when utilizing deterministic encoding instead of stochastic encoding is still capable of satisfying the secrecy and decodability constraints in the presence of training-phase jamming.  Hence, this subsection can be considered as the counterpart of Section~\ref{subsection:nowyner}. There, we assumed no training phase jamming, whereas here we mitigate training-phase jamming by other means.

In order to satisfy the security constraints without using stochastic encoding, the BS employs $\delta$-conjugate beamforming given in~\eqref{re_state_beamforming} and the strategy explained in Section~\ref{sub_section:counter}.
Specifically, Theorem~\ref{no_wyner2} and Corollary~\ref{revisit1} provide the  number of antennas that the BS requires in order to satisfy only the secrecy constraint and both the secrecy and decodability constraints, respectively. Note that Theorem~\ref{no_wyner2} and Corollary~\ref{revisit1} are the counterparts of Theorem~\ref{no_wyner} and Corollary~\ref{revisit}.
\begin{theorem}\textbf{(Establishing secrecy with no stochastic encoding)}\label{no_wyner2}  Let $\delta$, $\gamma>0$, and $\gamma+\delta>1$. Let block length be $T$ and length of data transmission phase be $T_d$. 
In the presence of  training-phase jamming,  for any $\epsilon>0$ and any rate tuple  $R\eqdef\left[R_1,\dots, R_K\right]$,  if $M\geq S_1(\epsilon)$ and $T_r\geq \max (M^{\gamma},K)$, then any deterministic code $\left(2^{BTR_1},\dots, 2^{BTR_K}, BT_d\right)$  employing $\delta$-conjugate beamforming satisfies
\allowdisplaybreaks
\begin{align}
\frac{1}{BT} H\left(W_k| Z^{BT_d}, H^B,\hat{H}^B,H^B_e \right) \geq R_k-\epsilon \label{norandom1}
\end{align}
for any $J\in\{1,\ldots, T_r\}$, $B\geq 1$, and $k\in[1:K]$, where  $$S_1(\epsilon)\eqdef \left(\frac{\rho_{max}M_e\max\left(1,\frac{\rho_{jam}}{\rho_r}\right)}{2^{\frac{T}{T_d}\epsilon}-1}\right)^{\frac{1}{\min(\delta,\delta+\gamma-1)}}$$and $\rho_{max}\eqdef \max_{k\in \{1,\dots,K\}}\rho_k$.\qed
\end{theorem}
The proof of Theorem~\ref{no_wyner2} can be found in Appendix~\ref{proof_of_no_wyner2}. Note that when $\gamma=1$ and $1\geq \frac{\rho_{\text{jam}}}{\rho_r}$, the necessary number of antennas to meet the secrecy constraint under training phase attack becomes identical to that under no attack. This result demonstrates the effectiveness of the defense strategy, \emph{hiding the pilot signal assignments from the adversary and expanding the pilot signal set}. 

There is a tradeoff between the number, $M$, of antennas and the length, $T_r$, of the training period necessary to satisfy constraints $M\geq S_1(\epsilon)$ and $T_r\geq \max (M^{\gamma},K)$ . This tradeoff is controlled by parameter $\gamma$. While choosing $\gamma$  close to $1$ minimizes $S_1(\epsilon)$ for any $\epsilon>0$, it increases the length of the training period, i.e., the overhead. To observe this: First, $S_1(\epsilon)$ is minimum at $\gamma =1$ due to the fact that  $\min(\delta,\delta+\gamma-1) \leq \delta$ and equality occurs when $\gamma =1$. Second, increasing $\gamma$ to $1$ also increases training overhead as $T_r$ has to be larger than $S_1(\epsilon)^\gamma$.


In Theorem~\ref{no_wyner2}, we provide the number of antennas required to satisfy only the secrecy constraint. Next corollary presents the number of antennas that BS needs in order to satisfy both the secrecy and the decodability constraints without need for stochastic encoding.

\begin{corollary}\textbf{(Any rate tuple is achievable with no need for stochastic encoding)}\label{revisit1}
 Let $0<\delta<1$, $\gamma+\delta>1$.  Let block length be $T$ and length of data transmission phase be $T_d$ and $J$ be any integer in $\{1,\dots,T_r\}$. 
In the presence of  training-phase jamming,  for any $\epsilon>0$ and any rate tuple  $R\eqdef\left[R_1,\dots, R_K\right]$,  if $M\geq \max\left(V_1(R),S_1(\epsilon)\right)$ and $T_r\geq M^{\gamma}$, then there exists  $B(\epsilon)>0$ and a sequence of  codes  $\left(2^{BT R_1}, \dots, 2^{BT R_K}, BT_d\right), B\geq B(\epsilon)$ that satisfy the constraints in~\eqref{cond1} and \eqref{l3} \textbf{without a need for stochastic encoding}, where  
\begin{align}
&V_1(R) \eqdef \max_{k\in \{1,\dots,K\}}\nonumber\\
&\left(\left(2^{R_k\frac{T}{T_d}}-1\right)\times \frac{\left(\rho_f+\rho_{jam}+1\right)\times \left(\rho_r+\rho_{jam}+1\right)}{\rho_r\rho_k}\right)^{\frac{1}{1-\delta}}.\nonumber
\end{align} \qed
\end{corollary}
The proof of Corollary~\ref{revisit1} can be found in Appendix~\ref{proof_of_revisit1}.
\subsection{How do we hide the pilot signal assignments?}
\label{subsection:share_securebits}
So far, we have demonstrated that, if pilot signal assignments can be kept secret from the adversary, the impact of training-phase jamming can be mitigated by increasing the cardinality of the pilot signal set at the expense of some increase in training overhead. Next, we discuss how to keep the assignments secret from the adversary.

In order to communicate the pilot signal assignments securely, at the beginning of each block, the BS shares with each user a secret key of size $\log L$ bits, that is unknown to the adversary. In the literature, by far the most popular way to generate an information-theoretically secure secret key across a wireless channels is via the use of reciprocal channel gains~\cite{wilson2007,mandayam2010,lai}. However, we cannot use such channel-gain based methods, since for those methods we need to observe the channel gains. However, our objective of generating the keys is to secure the training phase, whose sole purpose is to observe the channel gains in the first place, leaving us with a ``chicken or the egg'' dilemma.

With this observation, let us consider the methods in which these keys are generated and shared by standard private key based methods (e.g., Diffie-Hellman~\cite{diffie-hellman}) or public key based methods (e.g., RSA~\cite{rsa}). Thus, it only relies on existing standard computational cryptographic techniques and does not rely on information-theoretic techniques for secure key sharing. Note that a shared key between the BS and a user is used to encrypt the pilot signal assigned to that user and the encrypted assignment is communicated to the users immediately after key sharing. 

Despite the use of computational cryptographic methods for key generation, the security we provide has the ``same flavor'' as information theoretic secrecy, as we clarify next. The main drawback of computational cryptographic methods such as Diffie-Hellman is that, they make assumptions on the computational power of the adversaries. This kind of security is based on the supposition that, given that the key is hidden from an adversary via a difficult puzzle\footnote{For example, RSA is based on an NP problem: prime factorization of a large number.}, it takes an unreasonable amount of time for an adversary to crack it. Nevertheless, given enough time, the adversary will eventually decrypt the message (possibly quickly, given a quantum computer, for instance). This constitutes the main motivation for information-theoretic security, which makes no assumptions on the computational powers of the attackers.

In our approach, we have a hybrid scheme, combining information theoretic security and computational cryptography. We are using cryptography to hide the pilot sequence assignments, {\bf not the message}. Encrypting the pilot signal assignments is fundamentally different from encrypting the message. In message encryption, the signal received by the adversary remains vulnerable to cryptanalysis, long after the message is transmitted. On the other hand, with pilot signal assignment encryption, this window of time for cryptanalysis can be arbitrarily small: unless the adversary figures out the pilot sequence assigned to the targeted user before the training phase starts, the knowledge of the assignment becomes useless. But, we know that the training phase starts immediately after the encrypted assignment is communicated to the users. If we define the computational power required for the adversary as the ratio of amount of computation needed to decrypt the key via cryptanalysis to the time required to solve the problem, the computational power necessary for the adversary to make a damage on the targeted user goes to infinity. This addresses the shortcoming of existing cryptographic methods due to their assumptions on computational powers of adversaries. Note that, if the adversary cannot act during the training phase, the message transmission is ``perfectly secure'' as shown in Theorem 4.

It is important to emphasize that, in the above discussion, we did \textbf{not} show that the aforementioned defense strategy achieves information-theoretic security. Instead, we argued that, utilizing our defense strategy of encrypting training signals, we can avoid one of the main drawbacks of the existing computational-cryptographic methods, i.e., assumptions on the computational power of adversaries.

\section{Conclusion}
In this work, we study the physical-layer security of massive MIMO downlink communication. We first consider \emph{no training-phase jamming} attack in which the adversary jams \emph{only} the data communication and eavesdrops \emph{both} the data communication and training. We show that secure $DoF$ attained in the presence of no training-phase jamming is as same as the $DoF$ attained under no attack. This result shows the resilience of the massive MIMO  against adversaries not jamming the training phase. Further, we propose a joint power allocation and beamforming strategy, called $\delta$-conjugate beamfoming, using which we can establish information theoretic security without even a need for Wyner encoding as long as the number of antennas is above a certain threshold, evaluated in the sequel.

We next show the vulnerability of massive MIMO systems against the attack, called training-phase jamming in which the adversary jams and eavesdrops \emph{both} the training and data communication. We show that the maximum secure $DoF$ attained in the presence of training-phase jamming  is zero. We then develop a defense strategy against training-phase jamming. We show that if the BS keeps the pilot signal assignments hidden from the adversary and extends the cardinality of the pilot signal set, a secure $DoF$  equal to the maximum $DoF$ attained under no attack can be achieved. We finally provide a discussion why standard computational-cryptographic key sharing methods can be considered as strong candidates to encrypt the pilot signal assignments and how they achieve a level of security that is comparable to information-theoretically secure key-generation methods.

\appendices
\section{Proof of Theorem~\ref{thm:classical}} \label{app:classical_attack}
We first evaluate an upper bound on the secure DoF. In order to derive an upper bound, we assume that there is a single user and no adversary in the system. Further, we assume that the user and the BS have a perfect information of the channel gains. As a last assumption, the user is assumed to know the received pilot signals at the BS.  Hence, with these assumptions, the communication model in Section 2 reduces to a multiple input single output (MISO) communication set-up in which the channel gains and pilot signals are available at the BS and the user. Note that the capacity, the supremum of the achievable rates, of this new set-up upper bounds the secrecy rates achieved under the communication set-up explained in Section~\ref{system_model}. 
We derive the capacity with the following analysis:
\begin{align}
C = &\max_{p\left(x^{T_d}|h_1,y^{T_r}\right), \mathbb E\left[tr\left(X^{T_d}X^{T_d*}\right)\right]\leq \rho_f T_d}\nonumber\\
&\qquad\qquad\qquad\qquad\qquad\frac{1}{T} I\left(X^{T_d}; Y_1^{T_d}|Y^{T_r},H_1\right)\label{capacity}\\
&=\max_{p\left(x^{T_d}|h_1\right), E\left[tr\left(X^{T_d}X^{T_d*}\right)\right]\leq \rho_f T_d}\nonumber\\
&\qquad\qquad\qquad\quad\qquad\qquad\quad \frac{1}{T} I\left(X^{T_d}; Y_1^{T_d}|H_1\right)\label{capacity2}\\
&=\max_{p\left(x|h_1\right), \mathbb E\left[tr\left(XX^{*}\right)\right]\leq \rho_f} \frac{T_d}{T} I(X;Y|H_1)\label{capacity3}\\
&=\max_{\mathbb E\left[P(H_1)\right]\leq \rho_f}\frac{T_d}{T} \mathbb E\left[ \log\left(1+P\left(H_1\right)||H_1||^2\right)\right] \label{capacity4}
\end{align}
where $X^{T_d}$ is a complex $T_d\times M$  matrix and $P(\cdot): \mathbb C^{M}\to \mathbb R^{+} \cup \{0\}$ is a power allocation function. In the derivation above, \eqref{capacity} follows from Section 7.4.1 of \cite{gamal2012} where the capacity of a communication system in which the channel gains are available at both encoder and decoder is stated.  The equality in \eqref{capacity2} follows from the fact that $Y_1^{T_r}\to X^{T_d}, H_1\to Y^{T_d}_1$ forms a Markov chain and the equality in~\eqref{capacity3} follows from the fact that 
\begin{align} 
I\left(X^{T_d}; Y_1^{T_d}|H_1\right)&\leq \sum_{i=1}^{T_d} I(X_i;Y_i|H_1)\label{capacity_inequality}
\end{align}
and from the fact that the equality is attained in \eqref{capacity_inequality} if $p_{X^{T_d}|H_1}\left(x^{T_d}|h_1\right)= \prod_{i=1}^{T_d} p_{X|H_1}(x_i|h_1)$. Then, the RHS and the LHS of \eqref{capacity_inequality} becomes $I(X;Y|H_1)$.

In~\eqref{capacity4}, the equality follows from Section of \cite{tse2005}, where the capacity of MISO system is evaluated. In~\cite{tse2005}, the power allocation function maximizing~\eqref{capacity4} is given as 
\begin{equation}
  P\left(h_1\right) = \left(\lambda_M-\frac{1}{||h_1||^2}\right)^+, \nonumber
\end{equation}
where $\lambda_M$ is a non-negative real number and is chosen such that $\mathbb E[P(H_1)]=\rho_f$.  We next find an upper bound on $\lambda_M$ with the following analysis:
\begin{align}
\rho_f &=\mathbb E\left[\left(\lambda_M-\frac{1}{||H_1||^2}\right)^+\right]\nonumber\\
&\geq \lambda_M -\mathbb E \left[\frac{1}{||H_1||^2}\right] \label{gamma}\\
& = \lambda_M - \frac{1}{M-1} \label{gamma_bound}
\end{align}
where \eqref{gamma} follows from the fact that $\frac{1}{||H_1||^2}$ is distributed with inverse Gamma distribution and has a mean of $\frac{1}{M-1}$. Hence, we have $\lambda_M \leq \frac{1}{M-1}+\rho_f$ for $M> 1$. We next bound the $DoF$ of the MISO communication system as
\begin{align}
&\lim_{M\to \infty}\frac{T_d}{T} \frac{E\left[ \log\left(1+P\left(H_1\right)||H_1||^2\right)\right]}{\log M}\nonumber\\
&\leq \lim_{M\to \infty}\frac{T_d}{T} \frac{E\left[ \log\left(1+\lambda_M||H_1||^2\right)\right]}{\log M} \label{upp1}\\
&\leq \lim_{M\to \infty}\frac{T_d}{T}  \frac{\log\left(1+\lambda_ME\left[||H_1||^2\right]\right)}{\log M}\label{upp2}\\
&= \lim_{M\to \infty}  \frac{T_d}{T}\frac{\log\left(1+\lambda_MM\right)}{\log M}\nonumber\\
&\leq \lim_{M\to \infty}\frac{T_d}{T}  \frac{\log\left(1+\frac{M}{M-1}+M\rho_f\right)}{\log M}\label{upp3}\\
&=\frac{T_d}{T}+\lim_{M\to \infty}\frac{T_d}{T}  \frac{\log\left(\frac{1}{M}+\frac{1}{M-1}+\rho_f\right)}{\log M}\nonumber\\
&=\frac{T_d}{T}, \label{upp4}
\end{align}
where \eqref{upp1} follows from the fact that $P\left(\cdot\right)\leq \lambda_M$ for all realizations of $H_1$, \eqref{upp2} follows from Jensen's inequality, and \eqref{upp3} follows from~\eqref{gamma_bound}. In~\eqref{upp4}, we show that secure $DoF$ can be at most $\frac{T_d}{T}$. \qed

Next, we describe an achievability strategy to attain secure $DoF$ of $\frac{T_d}{T}$.
\\

\textbf{Channel estimation}: Pilot signals are mutually orthogonal, i.e.,
\begin{equation}
\phi_k \times \phi_l^{*}=
\begin{cases} 
T_r \rho_r&\mbox{if } k = l \\
0& \mbox{if }k\neq l
\end{cases}\nonumber
\end{equation} 
for any $k,l \in \{1,\ldots, K\}$. The BS employs MMSE for channel estimation. The estimated gain of the channel connecting the BS to $k$-th user is as follows:
\begin{equation}
\hat H_k = a H_k + b V_k\label{estimated_gain}
\end{equation}
for $k\in\{1,\ldots, K\}$, where $a\eqdef \frac{\rho_rT_r}{\rho_rT_r+1}$, $b\eqdef \frac{\sqrt{\rho_rT_r}}{\rho_rT_r+1}$, and $V_k$ is additive Gaussian noise distributed with $\mathcal{C}\mathcal{N}\left(0, I_{M}\right)$. Note that $\mathbb E\left[\hat H_k\right] = 0_{1\times M}$, $\mathbb E\left[\vert\vert \hat H_k\vert\vert^2\right]= Ma$.  Further,  for any $k\in [1:K]$ and for any $m,n\in \left[1:M\right]$,
$\mathbb E\left[\left\vert \hat H^*_{k_n} H_{k_m}\right\vert^2\right]=a^2+a$  if $m=n$, otherwise; $\mathbb E\left[\left\vert \hat H^*_{k_m} H_{k_n}\right\vert^2\right]=a^2$.\\

\textbf{Codebook generation:} Pick $R_k = \frac{T_d}{T} \log\left(1+\frac{M\rho_ka}{\rho_f+\rho_j +1}\right)-\frac{T_d}{T}\log\left(1+M_e\rho_k \right)$ and $\hat R_{k} = \frac{T_d}{T} \log\left(1+\frac{M\rho_ka}{\rho_f+\rho_j +1}\right)-\epsilon_1$ for some $\epsilon_1>0$ and for $k =1,\dots, K$. Generate $K$ codebooks, $c_k$, $k=1,\dots,K$, where $K$ is the number of users. Codebook $c_k$ contains independently and identically generated codewords, $s_{kl}^{BT_d}$, $l\in\{1,\ldots,2^{BT\hat R_{k}}\}$, each is drawn from $\mathcal C\mathcal N(\mathbf{0}, \rho_k I_{BT_d})$.

\textbf{Encoding:} In order to send $k$-th user's message $w_k\in \mathcal{W}_k$, the encoder draws index $l_k$ from the uniform distribution that has a sample space of 
$\left\{\left(w_k-1\right)2^{BT\left(\hat R_k-R_k\right)}+1,\ldots,w_k2^{BT\left(\hat R_k-R_k\right)}\right\}$. Note that this mapping makes the encoder \emph{stochastic}. The encoder then maps index $l_k$ to the corresponding codeword $s_{kl_k}^{BT_d}$ in codebook $c_k$.

The encoder employs a conjugate beamforming to map codewords to channel input sequence $X^{BT_d}$. The channel input at $j$-th channel use of $i$-th block can be written as follows:
\begin{equation}
X(i,j) = \sum_{k=1}^K   s_{kl_k} (i,j) \frac{1}{\sqrt{M \alpha_k}}\hat H^*_k(i)\nonumber
\end{equation}
where $\alpha_k = a$ for all $k\in\{1,\ldots, K\}$ due to the fact that  $\mathbb E\left[|\hat H_{k_m}|^2\right] = a$ for all $k\in\{1,\ldots, K\}$.

\textbf{Decoding}
Each user employs typical set decoding~\cite{cover1991}. Let $y_k^{BT_d}$ be the received signal at $k$-th user over $BT_d$ channel uses. The decoder at $k$-th user looks for an unique index $l_k\in\left\{1,\ldots,2^{BT_dR_{k}}\right\}$ such that $\left(s_{kl_k}^{BT_d}, y_k^{BT_d}\right) \in  \mathcal A^{BT_d}_{\epsilon}\left(S_k^{T_{d}}, Y_k^{T_d}\right)$, where $\mathcal A^{BT_d}_{\epsilon}\left(S_k^{T_{d}}, Y_k^{T_d}\right)$ is the set of jointly typical sequences ($s_k^{BT_d}, y_k^{BT_d}$) with 
\begin{align}
&Y_k^{T_d} =\frac{1}{\sqrt{Ma}}H_k \hat H_k^{*} S^{T_d}_k\nonumber\\
&\quad\qquad+\frac{1}{\sqrt{Ma}} \sum_{j=1, j\neq k}^K  H_j \hat H_j^{*} S^{T_d}_j +H_{jam,k} V_{jam}+ V_k\nonumber
\end{align}
where $S^{T_d}_j$ is distributed with $\mathcal{C}\mathcal{N}\left(0,\rho_k I_{T_d}\right)$, $j=1,\dots, K$ and $V_k$ is distributed with $\mathcal{C}\mathcal{N}\left(\mathbf{0},I_{T_d}\right)$.

\textbf{Probability error and equivocation analysis}
By the channel coding theorem~\cite{cover1991}, $\mathbb E\left[P_e\right]\to 0$ as $B \to \infty$ if $\hat R_k< \frac{T_d}{T}I\left(S_k^{T_d}, Y_k^{T_d}\right)$, $k=1,\dots, K$, where expectation is over random codebooks, $\mathcal C_1,\dots, \mathcal C_K$. Note that codebook $c_k$ is the realization of $\mathcal C_k$. Define 
\begin{align}
&T_0\eqdef  \frac{1}{\sqrt{Ma}}S_k \mathbb E\left[H_k \hat H_k^*\right]\nonumber\\ 
&T_1\eqdef \frac{1}{\sqrt{Ma}}S_k\left(E\left[H_k \hat H_k^*\right]-H_k \hat H_k^{*}\right)\nonumber\\
&T_2\eqdef \frac{1}{\sqrt{Ma}}\sum_{j=1, j\neq k}^K  H_k \hat H_j^{*} S_j\nonumber\\
& T_3 \eqdef H_{jam,k} V_{jam} +V_k.\nonumber
\end{align}
Note that $\mathbb E\left[T_0\right]= \mathbb E\left[T_1\right]=\mathbb E\left[T_2\right]=\mathbb E\left[T_3\right]=0$ and $\mathbb E\left[T_0 T_1^*\right]= \mathbb E\left[T_0T_2^*\right]=\mathbb E\left[T_0T_3^*\right]=0$.  We can bound $\frac{T_d}{T}I\left(S_k^{T_d}, Y_k^{T_d}\right)$ as 
\begin{align}
&\frac{T_d}{T}I\left(S_k^{T_d}, Y_k^{T_d}\right)\nonumber\\
&\geq \frac{T_d}{T} \log\left(1+\frac{\mathbb Var \left[T_0\right]}{\mathbb Var \left[T_1+T_2+T_3\right]} \right)\label{hassibi}\\
&=\frac{T_d}{T} \log\left(1+\frac{\mathbb Var \left[T_0\right]}{\mathbb Var \left[T_1\right]+\mathbb Var \left[T_2\right]+\mathbb Var \left[T_3\right]} \right)\label{cov}\\
& = \frac{T_d}{T} \log\left(1+\frac{M\rho_ka}{\rho_f + \rho_{jam}+1} \right), \label{varr1}
\end{align}
where \eqref{hassibi} follows from Theorem 1 of \cite{babak2000} and \eqref{cov} follows from the fact that $T_1$, $T_2$, and $T_3$ are uncorrelated random variables. The equality in~\eqref{varr1} follows from the fact that $\mathbb Var \left[T_0\right]=M\rho_ka$, $\mathbb Var \left[T_1\right]=\rho_k$, $\mathbb Var \left[T_2\right]= \sum_{j\neq k}\rho_j$, and $\mathbb Var\left[T_3\right] = \rho_{jam}+1$. From \eqref{varr1}, we conclude that $\hat R_k\leq \frac{T_d}{T}I\left(S_k^{T_d}, Y_k^{T_d}\right)$. Hence, $\mathbb E\left[P_e\right]\to 0$ as $B \to \infty$.  

We next analyze the secrecy constraint in~\eqref{l3}. Let $\left.H\left (W_k\right\vert Z^{TB},H^B,\hat{H}^B,H^B_e,\mathcal{C}\right)$ be  the  expectation of the conditional entropy in~\eqref{l3} over random codebooks $\mathcal C \eqdef \left[\mathcal C_1,\dots, \mathcal C_K\right]$. We show that the expectation satisfies the constraint in~\eqref{l3} for $k$-th user with the following analysis:\begin{align}
&\left.H\left (W_k\right\vert Z^{BT},G^B,\mathcal{C}\right)
\geq \left.H\left (W_k\right\vert Z^{BT},S^{BT_d},G^B,\mathcal{C}\right)\nonumber\\
&= \left.H\left (W_k\right\vert Z^{BT_d},S^{BT_d},G^B,\mathcal{C}\right)\label{training222}\\
&= \left.H\left(W_k,S_k^{BT_d}\right\vert Z^{BT_d},S^{BT_d}, G^B,\mathcal{C}\right)\nonumber\\
&\quad\qquad-\left.H\left(S_k^{BT_d}\right\vert W_k,Z^{BT_d},S^{BT_d}, G^B,\mathcal{C}\right)\nonumber\\
&\geq \left.H\left(S_k^{BT_d}\right\vert Z^{BT_d},S^{BT_d},G^B,\mathcal{C}\right)\nonumber\\
&\quad\qquad-\left.H\left(S_k^{BT_d}\right\vert W_k,Z^{TB},S^{BT_d}, G^B,\mathcal{C}\right)\nonumber\\
&=\left.H\left (S^{BT_d}_k\right\vert S^{BT_d},G^B,\mathcal{C}\right)\nonumber\\
&\qquad\qquad -\left.I\left (S^{BT_d}_k;Z^{BT_d}\right\vert S^{BT_d},G^B,\mathcal{C}\right)\nonumber\\
&\qquad\qquad-\left.H\left(S_k^{BT_d}\right\vert W_k,Z^{BT_d},S^{BT_d}, G^B,\mathcal{C}\right)\nonumber\\
&=\left.H\left (S^{BT_d}_k\right\vert\mathcal{C}_k\right)-\left.I\left (S^{BT_d}_k;Z^{BT_d}\right\vert S^{BT_d},G^B,\mathcal{C}\right)\nonumber\\
&\quad\qquad\qquad-\left.H\left(S_k^{BT_d}\right\vert W_k,Z^{BT_d},S^{BT_d}, G^B,\mathcal{C}\right)\label{indep1}\\
&=BT\hat R_k-\left.I\left (S^{BT_d}_k;Z^{BT_d}\right\vert S^{BT_d},G^B,\mathcal{C}\right)\nonumber\\
&\qquad\qquad\quad-\left.H\left(S_k^{BT_d}\right\vert W_k,Z^{BT_d},S^{BT_d}, G^B,\mathcal{C}\right)\label{uniform}\\
&\geq BT\hat R_k-\left.I\left (S^{BT_d}_k,\mathcal{C};Z^{BT_d}\right\vert S^{BT_d},G^B\right)\nonumber\\
&\qquad\qquad\quad-\left.H\left(S_k^{BT_d}\right\vert W_k,Z^{BT_d},S^{BT_d}, G^B,\mathcal{C}\right)\label{cont1}
\end{align}
where $G^B\eqdef \left[H^B, \hat H^B, H_e^B\right]$. Signal set  $S^{BT_d}\eqdef\left\{S^{BT_d}_i\right\}_{i\neq k}$ is defined to be the transmitted codewords of  the users except $k$-th user. Signals $Z^{BT_r}$ and $Z^{BT_d}$ are the received signals at the adversary over the training phases and data communication phases, respectively. Note that $Z^{BT}\eqdef\left[Z^{BT_r},\; Z^{BT_d}\right]$.

 In the above derivation~\eqref{training222} follows from the fact that  $Z^{BT_r}$ and  $(G^B, W_k,S^{BT_d},Z^{BT_d},\mathcal{C})$  are independent, \eqref{indep1} follows from the fact that $(S_k^{BT_d},\mathcal C_k)$ are independent with $(G^B, \left\{\mathcal C_i \right\}_{i\neq k})$, and~\eqref{uniform} follows from the fact that
$S_k^{BT_d}$ is uniformly distributed on a set of size $2^{BT\hat R_k}$. We continue the derivation as
\begin{align}
&\eqref{cont1}= BT\hat R_k-\left.I\left (S^{BT_d}_k;Z^{BT_d}\right\vert S^{BT_d},G^B\right)\nonumber\\
&\qquad\qquad\quad-\left.H\left(S_k^{BT_d}\right\vert W_k,Z^{BT_d},S^{BT_d}, G^B,\mathcal{C}\right)\label{markov1}\\
&\geq BT\hat R_k\nonumber\\
&\quad-\sum_{i=1}^B\sum_{j=T_r+1}^T\left.I\left (S_k(i,j);Z(i,j)\right\vert S(i,j),G(i)\right)\nonumber\\
&\quad\qquad\qquad-\left.H\left(S_k^{BT_d}\right\vert W_k,Z^{BT_d},S^{BT_d},G^B,\mathcal{C}\right)\nonumber\\
&\geq BT\hat R_k\nonumber\\
&\qquad- \sum_{i=1}^B \sum_{j=T_r+1}^T\mathbb E\left[\log\left(1+\frac{\rho_k}{Ma}\sum_{m=1}^{M_e}  \left\vert \hat H_kH^*_{e_m}\right\vert^2\right)\right]\nonumber\\
&\qquad\qquad\quad-\left.H\left(S_k^{BT_d}\right\vert W_k,Z^{BT},S^{BT_d}, G^B,\mathcal{C}\right)\label{gain_def_adv}\\
&\geq BT \hat R_k- BT_d\log\left(1+\frac{\rho_k}{Ma}\sum_{m=1}^{M_e}  \mathbb E\left[\left\vert \hat H_kH^*_{e_m}\right\vert^2\right]\right)\nonumber\\
&\qquad\qquad-\left.H\left(S_k^{BT_d}\right\vert W_k,Z^{BT},S^{BT_d}, G^B,\mathcal{C}\right)\nonumber\\
&= BT \hat R_k- B T_d\log\left(1+M_e\rho_k \right)\nonumber\\
&\qquad\qquad-\left.H\left(S_k^{BT_d}\right\vert W_k,Z^{BT},S^{BT_d}, G^B,\mathcal{C}\right)\label{evaluation1}\\
&\geq  BT\left(\hat R_k- \frac{T_d}{T}\log\left(1+M_e\rho_k \right)\right)- BT\epsilon_2\label{fano1}\\
& = BT\left(R_k-\epsilon\right)\label{equiconst}
\end{align}
for any $\epsilon_2>0$ and  sufficiently large $B$, where $\epsilon\eqdef \epsilon_1+\epsilon_2$ and $H_{e_m}$ in~\eqref{gain_def_adv} denotes the gain of the channel connecting 
$m$-th antenna at the BS to the adversary. The equality in \eqref{markov1} follows from the fact that  $\mathcal{C}\to S^{T_dB}_k, S^{T_dB},G^B\to Z^{T_dB}$ forms a Markov chain.  The equality in~\eqref{evaluation1} is due to the fact that $\mathbb E\left[\left\vert \hat H_k^*H_{e_m}\right\vert^2\right]=Ma$, $m=1,\dots, M_e$.

To get the inequality in~\eqref{fano1}, we need to bound $\frac{1}{BT}\left.H\left(S_k^{BT_d}\right\vert W_k,Z^{TB},S^{BT_d}, G^B,\mathcal{C}\right)$. Define $R_e \eqdef \hat R_k- R_k$. Note that $R_e < \frac{1}{T} \left.I\left(S_k^{T_d}; Z^{T_d}\right\vert S^{T_d}, G \right) =\frac{T_d}{T}\log\left(1+M_e\rho_k a\right)$. Hence, as in (52) of \cite{hybrid}, utilizing Fano's inequality and the channel coding theorem, we show that $\lim_{B\to\infty} \frac{1}{BT} \left.H\left(S_k^{BT_d}\right\vert W_k,Z^{TB},S^{BT_d}, G,\mathcal{C}\right) = 0$.

From the fact that $\mathbb E\left[P_e\right]\to 0$ as $B\to \infty$ and from \eqref{equiconst}, we conclude that there exists a sequence of codes satisfying constraints~\eqref{cond1} and ~\eqref{l3}. We now evaluate degree of freedom $d_k$ associated with $R_k$ as 
\begin{align}
d_k = \lim_{M\to\infty} \frac{R_k}{\log M} &= \frac{T_d}{T}+\lim_{M\to\infty} \frac{T_d}{T}\log\left(1+M_e\rho_k \right)\nonumber\\
& =\frac{T_d}{T}\nonumber
\end{align}
for $k=1,\dots, K$. Hence, the attained secure $DoF$ is equal to $\frac{T_d}{T}$.
\qed
\section{}
\subsection{Proof of Theorem~\ref{no_wyner}} \label{no_wyner_proof}
Note that since the adversary keeps silent during the training phases, the received signals at the BS over training phases are independent with $H_e^B$. Hence, we conclude that $\hat H^B\eqdef \left[ \hat H^B_1,\dots, \hat H^B_K\right]$ and $H_e^B$ are independent. 

The BS picks message rates $R_k>0$, $k=1,\ldots, K$. The equivocation rate for a code $\left(2^{BTR_1},\dots,2^{BTR_K}, BT_d\right)$ utilizing deterministic encoding mapping functions, $f_k$, $k=1,\ldots,K$ and $\delta$-conjugate beamforming is as follows:
\begin{align}
&\frac{1}{BT}\left.H\left (W_k\right\vert Z^{BT},G^B\right)\nonumber\\
&=\frac{1}{BT}\left.H\left (W_k\right\vert Z^{BT_d},G^B\right)\label{training11}\\
&\geq \frac{1}{BT}\left.H\left(W_k\right\vert Z^{BT_d},S^{BT_d}, G^B\right)\nonumber\\
&=\frac{1}{BT} \left.H\left(W_k\right\vert  S^{BT_d},G^B\right)\nonumber\\
&\qquad\qquad\qquad-\frac{1}{BT} \left.I\left (W_k;Z^{BT_d}\right\vert S^{BT_d},G^B\right)\nonumber\\
&= R_k -\frac{1}{BT} \left.I\left (W_k;Z^{BT_d}\right\vert S^{BT_d},G^B\right)\label{uniform11}\\
&\geq R_k -\frac{1}{BT} \left.I\left (S_k^{BT_d};Z^{BT_d}\right\vert S^{BT_d},G^B\right)\label{markov11}\\
&\geq R_k-\frac{1}{BT} \sum_{i=1}\sum_{j=1} \left.I\left (S_k(i,j);Z(i,j)\right\vert S(i,j),G(i)\right)\nonumber\\
&\geq R_k-\frac{1}{BT}\sum_{i=1}^B\sum_{j=T_r+1}^T\nonumber\\
& \qquad\quad\qquad\log\left(1+\frac{P_k(i,j)}{M^{1+\delta}\alpha_k}\sum_{m=1}^{M_e}  \mathbb E\left[\left\vert \hat H_k^*H_{e_m}\right\vert^2\right]\right)\label{pow_def11}\\
&\geq R_k-\frac{T_d}{T}\log\left(1+\frac{\rho_k}{M^{1+\delta}\alpha_k}\sum_{m=1}^{M_e}  \mathbb E\left[\left\vert \hat H_k^*H_{e_m}\right\vert^2\right]\right)\label{jensen11}\\
&=  R_k- \frac{T_d}{T}\log\left(1+\frac{M_e\rho_k}{M^{\delta}}\right)\label{independence11}\\
&\geq  R_k- \epsilon\nonumber
\end{align}
for any $\epsilon>0$ and for sufficiently large $M$, where $G\eqdef \left[H^B, \hat H^B, H_e^B\right]$. Particularly, for a given $\epsilon>0$ if $M \geq \left(\frac{M_e\rho_k}{2^{\frac{T}{T_d}\epsilon}-1}\right)^{\frac{1}{\delta}}$, then there exists a code that satisfies the constraint in~\eqref{cond3}.

In the above derivation, \eqref{training11} follows from the fact that $Z^{BT_r}$ and $(Z^{BT_d}, G^B,W_k)$ are independent, \eqref{uniform11} follows from the facts that $W_k$ is independent with $S^{BT_d}, G^B$ and uniformly distributed on $[1:2^{BTR_k}]$. In~\eqref{markov11}, the inequality follows from the fact that $W_k\to S_k^{BT_d}\to Z^{BT_d},S^{BT_d},G^B $.

In~\eqref{pow_def11}, $P_k(i,j)\eqdef \mathbb E\left||S_k(i,j)||^2\right]$, where the expectation is over $W_k$. In~\eqref{jensen11}, the inequality follows from Jensen's inequality and from the fact that $\frac{1}{BT_d}\sum_{i=1}^B\sum_{j=T_r+1}^{T} P_k(i,j) \leq \rho_k$. In~\eqref{independence11}, the equality follows from the fact $\hat H_k$ and $H_{e_m}$ are independent and $\mathbb E\left[\left\vert \hat H_k^*H_{e_m}\right\vert^2\right]= M\alpha_k$, $k=1,\dots, K$.

\subsection{Proof of Corollary~\ref{revisit}}\label{proof:revisit}
Pick  $0<\delta<1$. Pick arbitrary $\epsilon>0$ and rate tuple $R=[R_1, \dots,R_K]$. Let  $M\geq \max(V(R),S(\epsilon))$. Note that inequality $M> V(R)$ implies that  $R_k < \frac{T_d}{T} \log\left(1+\frac{M^{1-\delta}a\rho_k}{M^{-\delta}\rho_f+\rho_j +1}\right)$, $k=1,\dots, K$. We first show that there exists $B(\epsilon)>0$  and a sequence of  codes $\left(2^{BT R_1},\dots, 2^{BT R_K}, BT_d\right)$  utilizing $\delta$-beamforming and deterministic mapping, that satisfy the decodability constraint in~\eqref{cond1} for $B\geq B(\epsilon)$.

The same channel estimation strategy in Appendix A is used. Codebook generation is as same as the one in Appendix~\ref{app:classical_attack}. The BS generates $K$ codebooks, $c_k$, $k=1,\dots, K$, where $c_k$ contains $2^{BTR_k}$ codewords, $s_{kl}^{BT_d}$, $l\in\{1,\ldots,2^{BTR_k}\}$. 

To send $k$-th user's message $w_k\in \mathcal W_k=\left\{1,\ldots,2^{BTR_k}\right\}$, the BS maps message $w_k$ to the corresponding codeword $s_{kw_k}^{BT_d}$ in codebook $c_k$. Note that there is no randomization in the mapping as opposed to the mapping in the encoding in Appendix~\ref{app:classical_attack}, where the codeword is a stochastic function of the message. The BS employs $\delta$-conjugate beamforming to map codewords to channel input sequence $X^{BT_d}$. The channel input at $j$-th channel use of $i$-th block can be written as
\begin{equation}
X(i,j) = \sum_{k=1}^K   s_{kw_k} (i,j) \frac{1}{\sqrt{M^{1+\delta}\alpha_k}}\hat H^*_k(i)
\end{equation}
where $ \alpha_k = a$ and $a$ is defined in \eqref{estimated_gain}.

The typical set decoding is used at each user as in the proof of Theorem~\ref{thm:classical} in Appendix~\ref{app:classical_attack}. Hence, since $R_k< \frac{1}{T}I\left(S_k^{BT_d};Y_k^{BT_d}\right)$, $k=1,\dots,K$, by the channel coding theorem, there exists a sequences of codes that satisfy constraint~\eqref{cond1}.


In addition, since $M\geq S(\epsilon)$, the sequence of codes mentioned above satisfy the secrecy constraint in~\eqref{l3} due to Theorem~\ref{no_wyner}.  Hence, the proof of Corollary~\ref{revisit} follows.
\qed

\section{}
\subsection{Proof of Theorem~\ref{zerodof}} \label{proof:zerodof}
Throughout the proof, we assume that the BS employs conjugate beamforming without loss of generality. 
 Suppose that $R_k$ is an achievable rate. From the constraints~\eqref{cond1}-\eqref{l3} and Fano's inequality, we have
\begin{align}
&\frac{1}{BT} H\left(W_k| Z^{BT_d},H^B,\hat H^B, H^B_e\right)\geq R_k- \delta_{B}\label{sec444}\\
&\frac{1}{BT} \left.H\left(W_k\right\vert Y_k^{BT_d}\right) \leq \epsilon_{B} \label{rel444}
\end{align}
 where $\epsilon_{B}$ and $\delta_{B}$ go to zero as $B\to \infty$.  

The LHS of~\eqref{sec444} can be written as follows
\begin{align}
&\frac{1}{BT}H\left(W_k| Z^{BT_d},H^B,\hat H^B, H^B_e\right)\nonumber\\
&=\frac{1}{BT}H\left(W_k| Z^{BT_d},\hat H^B, H^B_e\right)\label{obs_eq2}\\
&=\frac{1}{BT}H\left(W_k| \tilde Z^{BT_d},\tilde H^B, H^B_k\right)\label{obs_eq3}\\
&=\frac{1}{BT}H\left(W_k| \tilde Z^{BT_d},\tilde H^B, H^B\right)\label{obs_eq4}
\end{align}
where $\tilde H(i)\eqdef \left[\hat H_1(i), \ldots, \tilde H_k(i),\ldots, \hat H_K(i)\right]$  and
\begin{align}
&\tilde Z(i,j) \eqdef \frac{1}{\sqrt{M\alpha_k}}H_k(i)\tilde H_k^*(i)S_k(i,j)\nonumber\\
&\qquad\qquad+\sum_{l=1,l\neq k}^K \frac{1}{\sqrt{M\alpha_l}}H_l(i)\tilde H_l^*(i)S_l(i,j)  +W
\end{align}
for $1\leq i\leq B$ and $T_d+1\leq j\leq T$.
The equality in~\eqref{obs_eq2} follows from the fact  that  $H^B\to Z^{BT_d}, \hat H^B, H^B_e\to W_k$ forms a Markov chain and the equality in~\eqref{obs_eq3} follows from the fact that the joint distribution of $W_k, Z^{BT_d},H^B_e, \hat H^B_1,\ldots, \hat H^B_k,\ldots \hat H^B_K$ is identical with that of $W_k, \tilde Z^{BT_d},H^B_k, \hat H^B_1, \ldots, \tilde H^B_k,\ldots, \hat  H^B_K$. The equality in~\eqref{obs_eq4} follows from the fact that  $H^B / H^B_k\to \tilde Z^{BT_d}, \hat H_1^B, \tilde H^B\to W_k$ forms a Markov chain.

The upper bound on $R_k$ can be derived with the following steps:
\begin{align}
&R_k\leq \frac{1}{BT}\left.H\left(W_k\right\vert Z^{BT_d},H^B,\hat H^B, H^B_e \right)\nonumber\\
&\qquad\qquad-\frac{1}{BT}\left.H\left(W_k\right\vert Y_k^{BT_d}\right)+\gamma_B\label{Fanos444}\\
&= \frac{1}{BT}\left.H\left(W_k\right\vert \tilde Z^{BT_d},\tilde H^B, H^B \right)\nonumber\\
&\qquad\qquad-\frac{1}{BT}\left.H\left(W_k\right\vert Y_k^{BT_d}\right)+\gamma_B\label{Fanos445}\\
&\leq \frac{1}{BT}\left.H\left(W_k\right\vert \tilde Z^{BT_d},\tilde H^B, H^B \right)\nonumber\\
&\quad-\frac{1}{BT}\left.H\left(W_k\right\vert \tilde Z^{BT_d},\tilde H^B, H^B\right)+\gamma_B\label{cond_reduc444}\\
&=\left.\frac{1}{BT}I\left(W_k; Y_k^{BT_d}\right\vert \tilde Z^{BT_d}, \tilde H^B,H^B \right)\nonumber\\
&\leq\left.\frac{1}{BT}I\left(S^{BT_d}; Y_k^{BT_d}\right\vert \tilde Z^{BT_d},G^B  \right)+\gamma_B\label{markov444}\\
&\leq \frac{1}{BT}\sum_{i=1}^B\sum_{j=T_r+1}^T\left.I\left(S(i,j); Y_k(i,j)\right\vert 
\tilde Z(i,j),G \right)+\gamma_B
\label{memoryless} \\
&= \int\frac{1}{BT}\sum_{i=1}^B\sum_{j=T_r+1}^T\left.I\left(S(i,j); Y_k(i,j)\right\vert Z(i,j),g \right)\nonumber\\
&\qquad\qquad\quad\qquad\quad\qquad\qquad\ p_{G}(g) \, \mathrm{d} g+\gamma_B,\label{con1444}
\end{align}
where $\gamma_B\eqdef \epsilon_B+\delta_B$, $G^B \eqdef \left[H^B,\tilde H^B\right]$, and $G\eqdef \left[H,\tilde H\right]$. In the derivation above, \eqref{Fanos444} follows from \eqref{sec444} and \eqref{rel444}, and \eqref{cond_reduc444} follows from the fact that conditioning reduces the entropy. The inequality in~\eqref{markov444} follows from the fact that $W_k\to S^{BT_d}\to Y_k^{BT_d}, \tilde Z^{T_d}, G^B$. The inequality in~\eqref{memoryless} follows from the memoryless property of the channel and from the assumption in Theorem~\ref{zerodof}, stating that  $\left(\tilde H(i), H(i)\right)$ have an identical probability distribution for any $i\geq1$. We continue the upper bound derivation with the following steps: 
\begin{align}
&\eqref{con1444}\leq \int\frac{1}{BT}\sum_{i=1}^B\sum_{j=T_r+1}^{T}\left.I\left(S_G(i,j); Y_k(i,j)\right\vert \tilde Z(i,j),g \right) \nonumber\\
&\qquad\qquad\qquad\qquad\qquad\qquad\qquad  p_{G}(g)\, \mathrm{d}g +\gamma_B\label{Gaussian_max}\\
&\leq\frac{T_d}{T}\int\left.I\left(S_G; Y_k\right\vert Z,g \right) p_{G}(g) \, \mathrm{d}g+\gamma_B \label{concavity444} \\
&\leq \frac{T_d}{T}\int \left[\max_{\Sigma \in \mathcal S} \left(\log\left(1+ c_k \Sigma c^{*}_k\right)-\right.\right.\nonumber\\
&\qquad\qquad\left.\left.\log\left(1+c_e\Sigma c^{*}_e\right)\right)\right]^{+} p_{G}(g) \,  \mathrm{d}g +\gamma_B\label{maximize444}\\
&\leq \frac{T_d}{T}\mathbb E \left[\max_{\Sigma \in \mathcal S} \left(\left[\log\left(1+ C_k \Sigma C^{*}_k\right)-\right.\right.\right.\nonumber\\
&\qquad\qquad\qquad\qquad\left.\left.\left.\log\left(1+C_e\Sigma C^{*}_e\right)\right]^+\right)\right] +\gamma_B,\label{maxiize444}
\end{align}
where $C_k$ and $C_e$ are $1\times K$ random vectors and are defined as  $C_k\eqdef \left[\frac{H_k\hat H_1^{*}}{\sqrt{M\alpha_1}},\dots,\frac{H_k\hat H^*_k}{\sqrt{M\alpha_k}},\ldots \frac{H_k\hat H_K^{*}}{\sqrt{M\alpha_K}}\right]$ and $C_e\eqdef \left[ \frac{H_k \hat H_1^{*}}{\sqrt{M\alpha_1}},\ldots,\frac{H_k\tilde H^*_k}{\sqrt{M\alpha_k}},\ldots,\frac{H_k \hat  H_K^{*}}{\sqrt{M\alpha_K}}\right]$. Further, $c_k$ and $c_e$ are  the realizations of $C_k$ and $C_e$, respectively. Define $\Sigma_{ij}$ as $K\times K$ covariance matrix of $S_k(i,j)$. Note that $\Sigma_{ij}$ is a diagonal matrix due to the fact that each component of $S(i,j)$ are independent. The inequality~\eqref{Gaussian_max} follows from ($41$) of \cite{babak2010}, where $S_G(i,j)$ in ~\eqref{Gaussian_max} is distributed with $\mathcal{C}\mathcal{N}\left(0_{K\times K},\Sigma_{ij}\right)$.

Define $f (\Sigma_{ij}) \eqdef \left.I\left(S_G(i,j); Y_k(i,j)\right\vert \tilde Z(i,j),g \right)$. The inequality in~\eqref{concavity444} follows from Jensen's inequality and 
Proposition 5 of~\cite{babak2010} that states $f(\Sigma_{ij})$ is a concave function of $\Sigma_{ij}$. Note that $S_G$ in~\eqref{concavity444} is distributed with $\mathcal C\mathcal N\left(0,\frac{1}{BT_d}\sum_{i=1}^{B}\sum_{j=T_r+1}^T\Sigma_{ij}\right)$. The inequality in~\eqref{maximize444} follows from (139) of \cite{babak2010}, where $\mathcal S$ is a set of covariance matrices and defined as 
\begin{align}
&\mathcal S \eqdef \left\{\Sigma: \Sigma  	\preceq diag\left(\rho_1,\dots,\rho_K\right)\right.\nonumber\\
&\qquad\qquad\qquad\qquad\left.\text{ and } \Sigma \text{ is a diagonal matrix} \right\} \label{constraint_matrix}
\end{align}
We can rewrite the random variable inside the expectation  as
\begin{align}
&\max_{\Sigma \in \mathcal S} \left(\left[\log\left(1+ \frac{1}{M}C_k \Sigma C^{*}_k\right)-\right.\right.\nonumber\\
&\qquad\qquad\qquad\qquad\qquad\left.\left.\log\left(1+\frac{1}{M}C_e\Sigma C^{*}_e\right)\right]^+\right)\label{reference444}\\
&=\max_{\Sigma \in \mathcal S}\left(\left[\log\left(\frac{1}{M}+ \rho_k(G)v_k(G)+\sum_{l\neq k}^K \rho_l(G)v_l(G)\right)\right.\right.\nonumber\\
&\left.\left. -\log\left(\frac{1}{M}+\rho_k(G)w_k(G)+ \sum_{l\neq k}^K \rho_l(G)v_l(G)\right)\right]^+\right) \label{reference1444}
\end{align}
with probability 1, where $\rho_k(G)$ is defined to be $k$-th element on the diagonal of $\Sigma$, i.e., $\Sigma \eqdef diag\left(\rho_1(G), \dots, \rho_K(G)\right)$. Note that 
\begin{equation}
0\leq \rho_l(G)\leq\rho_l, \;l=1,\dots,K,\nonumber
\end{equation}
due to \eqref{constraint_matrix}. In~\eqref{reference1444}, we define
\begin{align}
v_l(G)\eqdef \frac{1}{\alpha_lM^2} \left\vert H_k \hat H_l^{*}\right\vert^2\nonumber
\end{align}
for $l=1,\dots, K$ and $w_k(G)\eqdef   \frac{1}{\alpha_kM^2}\left\vert H_k \tilde H_k^{*}\right\vert^2$.
We continue to simplify \eqref{reference444} with the following:
\begin{align}
&\eqref{reference1444}= \left[\max_{\rho_k(G):0\leq \rho_k(G)\leq \rho_k}\left(\log\left(\frac{1}{M}+ \rho_k(G)v_k(G)\right)\right.\right.\nonumber\\
&\quad\qquad\qquad\qquad\left.\left. -\log\left(\frac{1}{M}+\rho_k(G)w_k(G) \right)\right)\right]^+\label{main_result1}\\
&=\left[\left(\log\left(\frac{1}{M}+ \rho_kv_k(G)\right)\right.\right.\nonumber\\
&\quad\qquad\qquad\qquad\left.\left. -\log\left(\frac{1}{M}+\rho_kw_k(G) \right)\right)\right]^+\label{main_result}
\end{align}
with probability $1$, where \eqref{main_result1} follows from the fact that $f(x)=\left[\log(a+x)-\log(b+x)\right]^+$ is a non-increasing  function if $x\geq 0$, where $a$ and $b$ are positive real numbers. The equality in~\eqref{main_result} follows from the fact $g(x)=\left[\log\left(\frac{1}{M}+ ax\right)-\log\left(\frac{1}{M}+bx\right)\right]^+$ is non-decreasing if $x\geq 0$ where $a$ and $b$ are non-negative real numbers and $M\geq 1$.

 We now bound $R_k$ as follows:
\begin{align}
&R_k\leq \frac{T_d}{T}\mathbb E\left[\left[\log\left(\frac{1}{M}+ \rho_k v_k(G)\right) \right.\right.\nonumber\\
&\qquad\left.\left. -\log\left(\frac{1}{M}+\rho_k w_k(G)\right)\right]^+\right]+\gamma_B \label{conc1_to_rate}\\
&=\frac{T_d}{T}\mathbb E\left[\left[\log\left(\frac{1}{M}+ \rho_k v_k(G)\right) \right.\right.\nonumber\\
&\qquad\left.\left. -\log\left(\frac{1}{M}+\rho_k w_k(G)\right)\right]^+\right] \label{conc2_to_rate}
\end{align}
where \eqref{conc1_to_rate} follows from \eqref{maxiize444} and from the fact that $\eqref{reference444}=\eqref{main_result}$ with probability 1 and \eqref{conc2_to_rate} follows from the fact that $\lim_{B\to\infty} \gamma_B = 0$.

We now bound the secure degree of freedom of $k$-th user as follows
\begin{align}
&d_k=\lim_{M\to\infty}\frac{R_k}{\log M}\nonumber\\
&\leq  \lim_{M\to\infty}\frac{T_d}{T}\mathbb E\left[\left[\frac{\log\left(1+M \rho_k v_k(G)\right)}{\log M} \right.\right.\nonumber\\
&\qquad\qquad\qquad\qquad\left.\left. -\frac{\log\left(1+M\rho_k w_k(G)\right)}{\log M}\right]^+\right]\label{main1}\\
&=  \frac{T_d}{T}\mathbb E\left[\lim_{M\to\infty}\left[\frac{\log\left(1+M \rho_k v_k(G)\right)}{\log M} \right.\right.\nonumber\\
&\qquad\qquad\qquad\qquad\left.\left. -\frac{\log\left(1+M\rho_k w_k(G)\right)}{\log M}\right]^+\right]\,\label{main2},
\end{align}
where  \eqref{main1} follows from \eqref{conc2_to_rate} and \eqref{main2} follows form the dominant convergence theorem. To apply the dominant convergence theorem, we need to show that random variable
\begin{align}
&t(M)\eqdef \left[\frac{\log\left(1+M \rho_kv_k(G)\right)}{\log M} \right.\nonumber\\
&\qquad\qquad\qquad\left. -\frac{\log\left(1+M\rho_k w_k(G)\right)}{\log M}\right]^+\label{dct1}
\end{align}
is upper and lower bounded by random variables that have a finite limit for $M>1$. Note that $t(M)$ is lower bounded by zero and upper bounded by 
\begin{align}
t^+(M)\eqdef \frac{\log\left(1+M\rho_kv_k(G)\right)}{\log M}\nonumber
\end{align}
for any $M>1$ since the second $\log(\cdot)$ term in~\eqref{dct1} is non-negative. We next upper bound $\mathbb E\left[t^{+}(M)\right]$ as follows:
\begin{align}
&\mathbb E\left[t^{+}(M)\right]\nonumber\\
&=\mathbb E\left[ \frac{\log\left(\frac{1}{M}+ \rho_k v_k(G)\right)}{\log M}\right]+1\label{dctfirst}\\
&\leq\mathbb E\left[ \log\left(\frac{1}{M}+ \rho_k v_k(G)\right)\right]+1\nonumber\\
&\leq \log\left(1+\rho_k  \mathbb E \left[v_k(G)\right]\right)+1\label{dct1_jensen}\\
&\leq\log\left(1+\rho_k\left(\gamma_k+\pi_k\right) \right)+1\label{dct1_secondorder}\\
&< \infty, \label{dctlast}
\end{align}
where $\gamma_k\eqdef \left\vert  \mathbb E\left[H_{k_m}\hat H^*_{k_m}\right]\right\vert^2$, $\pi_k\eqdef  \mathbb E\left[\left\vert H_{k_m}\hat H^*_{k_m}\right\vert^2\right]$. In the derivation above, \eqref{dct1_jensen} follows from Jensen's inequality and \eqref{dct1_secondorder} follows from the fact that 
$\mathbb E\left[\left\vert H_k\hat H_k^{*}\right\vert^2\right]=\left( M^2-M\right) \left\vert  \mathbb E\left[H_{k_m}\hat H^*_{k_m}\right]\right\vert^2 + M \mathbb E\left[\left\vert H_{k_m}\hat H^*_{k_m}\right\vert^2\right] =\left( M^2-M\right)\gamma_k+M\pi_k$.

We continue the derivation of the upper bound on $d_k$ with the following:
\begin{align}
&\eqref{main2}=   \frac{T_d}{T}\mathbb E\left[\left[\lim_{M\to\infty}\frac{\log\left(\frac{1}{M}+\rho_k v_k(G)\right)}{\log M} \right.\right.\nonumber\\
&\qquad\left.\left. -\lim_{M\to\infty}\frac{\log\left(\frac{1}{M}+\rho_k w_k(G)\right)}{\log M}\right]^+\right]\label{main3}\\
&=0, \label{main4}
\end{align}
where \eqref{main3} follows from the fact $\left[\cdot\right]$ is a continuous function. In order to show the equality in~\eqref{main4}, first note that
\begin{align}
&\lim_{M\to\infty}v_k(G)=\lim_{M\to\infty}\frac{1}{\alpha_kM^2} \left\vert H_k \hat H_k^{*}\right\vert^2\nonumber\\
&=\lim_{M\to\infty} \frac{1}{\alpha_kM}\sum_{m=1}^{M} H_{k_m} \hat H^{*}_{k_m}\times\nonumber\\
&\qquad\qquad\qquad\qquad \lim_{M\to\infty} \frac{1}{M}\sum_{m=1}^{M} H^{*}_{k_m} \hat H_{k_m}\nonumber\\
&=\frac{1}{\alpha_k}\left\vert\mathbb E \left[ H_{k_m}\hat H^*_{k_m}\right]\right\vert^2,\label{law1}
\end{align}
with probability $1$, where \eqref{law1} follows from the strong law of large numbers. In a similar way we can show that 
\begin{align}
&\lim_{M\to\infty}w_k(G)=\frac{1}{\alpha_k}\mathbb E \left[\left\vert  H_{k_m}\tilde H^*_{k_m} \right\vert^2\right]\nonumber\\
&=\frac{1}{\alpha_k}\left\vert\mathbb E \left[ H_{e_m}\hat H^*_{k_m} \right]\right\vert^2 \label{identical_dist}
\end{align}
with probability $1$, where~\eqref{identical_dist} follows from the fact that the joint probability distribution of $\left(H_e, \hat H_k\right)$ is identical with that of $\left(H_k, \tilde H_k\right)$. Hence, we have
\begin{align}
&\lim_{M\to\infty}\log\left(\frac{1}{M}+ \rho_k v_k(G)\right)=\log\left(\lim_{M\to\infty}\rho_k v_k(G)\right)\nonumber\\
&\quad\quad\qquad\qquad\qquad=\log\left(\frac{\rho_k}{\alpha_k} \left\vert\mathbb E \left[ H_{k_m}\hat H^*_{k_m} \right]\right\vert^2\right)\label{law33}
\end{align}
with probability $1$. Further, we have
\begin{align}
&\lim_{M\to\infty}\log\left(\frac{1}{M}+ \rho_k w_k(G)\right)=\log\left(\lim_{M\to\infty} \rho_k w_k(G)\right)\nonumber\\
&\quad\qquad\qquad\qquad\qquad=\log\left(\frac{\rho_k}{\alpha_k} \left\vert\mathbb E \left[ H_{e_m}\hat H^*_{k_m} \right]\right\vert^2\right)\label{law34}
\end{align}
with probability 1. The equality in~\eqref{main4} follows by combining~\eqref{law33} and~\eqref{law34}. Hence, the proof ends.

The proof of Theorem~\ref{zerodof} for the case in which the BS employs $\delta$-conjugate beamforming can be done in the similar way. One only needs to  replace $c_k \Sigma c^{*}_k$ and $c_e \Sigma c^{*}_e$ in \eqref{maximize444} with $\frac{1}{M^\delta}c_k \Sigma c^{*}_k$ and $\frac{1}{M^{\delta}}c_e \Sigma c^{*}_e$, respectively and change the rest of the proof accordingly.

\qed
\subsection{Proof of Corollary~\ref{zerorate}} \label{proof:zerorate}
Assume that the BS employs conjugate beamforming without loss of generality. Note that from~\eqref{conc2_to_rate}, we have following upper bound:
\begin{align}
&\lim_{M\to\infty}R_k=\lim_{M\to\infty}\frac{T_d}{T}\mathbb E\left[\left[\log\left(\frac{1}{M}+\rho_k v_k(G)\right) \right.\right.\nonumber\\
&\qquad\left.\left. -\log\left(\frac{1}{M}+\rho_k w_k(G)\right)\right]^+\right] \nonumber\\
&=\frac{T_d}{T}\mathbb E\left[\lim_{M\to\infty}\left[\log\left(\frac{1}{M}+ \rho_k v_k(G)\right) \right.\right.\nonumber\\
&\qquad\qquad\left.\left. -\log\left(\frac{1}{M}+\rho_k w_k(G)\right)\right]^+\right] \label{conc4_to_rate}
\end{align}
where~\eqref{conc4_to_rate} follows from the dominant convergence theorem.  To apply the dominant convergence theorem, we need to show that random variable
\begin{align}
&g(M)\eqdef \left[\log\left(\frac{1}{M}+ \rho_k v_k(G)\right) \right.\nonumber\\
&\qquad\left. -\log\left(\frac{1}{M}+\rho_k w_k(G)\right)\right]^+\nonumber
\end{align}
is upper and lower bounded by random variables that have a finite limit for $M>1$. Note that $g(M)$ is lower bounded by zero and upper bounded by 
\begin{align}
&g(M)\leq \left[\log\left(\frac{1}{M}+ \rho_k v_k(G)\right) \right.\nonumber\\
&\qquad\qquad\qquad\left. -\log\left(\frac{1}{M}+\rho_k w_k(G)\right)\right]^+\nonumber\\
&\leq \log\left(2+ \rho_k v_k(G)+\rho_kw_k(G)\right) -\log\left(\rho_k w_k(G)\right)\label{dct224}
\end{align}
with probability $1$ for any $M>1$. Noting the analysis in \eqref{dctfirst}-\eqref{dctlast}, in order to show \eqref{dct224} is upper bounded by a random variable that has a finite expectation, it is sufficient to show the expectation of second $\log(\cdot)$ term in~\eqref{dct224} has a finite lower bound. Hence,
\begin{align}
&\mathbb E\left[\log\left(\rho_k w_k(G)\right)\right]=\log \rho_k+\mathbb E\left[\log\left(w_k(G)\right)\right]\nonumber\\
&=\log \rho_k+\mathbb E\left[\log\left(w_k(G)\right)\right]\nonumber\\
&=\log \rho_k-\log \alpha_k+\mathbb E\left[\log\left(K_M\right)\right]\label{equivalent1}\\
&\geq \log \rho_k-\log \alpha_k+\int \log(x)p_{K_M}(x) \,  \mathrm{d}x\nonumber\\
&\geq \log \rho_k-\log \alpha_k+\int_0^1 \log(x) p_{K_M}(x) \,  \mathrm{d}x\nonumber\\
&\geq \log \rho_k-\log \alpha_k+\int_0^1 \log(x) r \,  \mathrm{d}x\label{assumption2}\\
&= \log \rho_k-\log \alpha_k-r \log e\nonumber\\
&> -\infty,\nonumber
\end{align}
where $r$ is defined in the statement of Corollary~\ref{zerorate}. In~\eqref{equivalent1}, the equality follows from the definition of $K_M$ in Corollary~\ref{zerorate} and from the fact that the joint probability distribution of $\left(H_e, \hat H_k\right)$ is identical with that of $\left(H_k, \tilde H_k\right)$. In~\eqref{assumption2}, the inequality follows from the assumption in Corollary~\ref{zerorate}. The rest of the proof follows from Appendix~\ref{proof:zerodof}

The proof for the case the BS employs $\delta$-conjugate beamforming follows from the same argument at the end of Appendix~\ref{proof:zerodof}\qed
\section{Proof of Theorem~\ref{encrypting}}\label{proof_of_encrypting}
The length, $T_r$  of training phase has to be identical to at least the size of the pilot signal set $L$ so that the BS can generate $L\geq K$ mutually orthogonal pilot signals. Let $J$ be any integer in set $\{1,\dots,T_r\}$. In order to estimate $k$-th user's channel, the BS first projects the received signal during the training phase $Y^{T_r}$ indicated in~\eqref{l333} to $\phi_k$. Then, the BS normalizes the projected signal and estimates the gain of the channel connecting the BS to $k$-th user at $i$-th block as
\begin{align}
&\hat H_{k}(i)=x_1\left(\sqrt{T_r\rho_r}H_{k}(i)\right.\nonumber\\
&\qquad\qquad\qquad\left.+\Pi_i\sum_{n=1}^{M_e} \sqrt{\frac{T_r\rho_{jam}}{M_eJ}}H_{e}(i) + V_k\right)
\label{chest}
\end{align}
for any $k\in \{1,\ldots,K\}$, where $\mathbb E\left[||\hat H_k(i)||^2\right]=1$, $V_k$ is distributed as $\mathcal{C}\mathcal{N}(0,I_{M})$ for any $k\in \{1,\ldots,K\}$, $x_1\eqdef \frac{1 }{\sqrt{M}\sqrt{T_r\rho_r  +1+T_r\frac{\rho_{jam}}{J}}}$ and $\{\Pi_i\}_{i\geq 1}$ is an i.i.d Bernoulli process, where $\mathbb P(\Pi_i =1) = \frac{J}{L}$. Event $\{\Pi_i=1\}$ indicates that the set of pilot signals the adversary contaminates at $i$-th block contains $k$-th user's pilot signal.

Utilizing stochastic encoding and conjugate beamforming as in the proof Theorem~\ref{thm:classical}, we can show that rate 
\begin{align}
R_k =&\left[\frac{T_d}{T} \log\left(1+\frac{M\rho_ka}{\rho_f+\rho_{jam} +1}\right)\right.\nonumber\\
&\quad\left.-\frac{T_d}{T}\log\left(1+M_e\rho_k+\frac{MM_e\rho_k\rho_{jam}a}{L\rho_r} \right)\right]^+\label{rate1111}
\end{align}
for any $k\in \{1,\dots,K\}$ is achievable, where $a\eqdef \frac{T_r\rho_r}{T_r\rho_r  +1+T_r\frac{\rho_{jam}}{L}}$. Notice that the rate in~\eqref{rate1111} does not depend on $J$.  We can rewrite  $R_k$ as
\begin{align} 
R_k =&\left[\frac{T_d}{T} \log\left(1+\frac{M\rho_k \rho_r T_r}{(\rho_f+\rho_{jam} +1)(\rho_rT_r+\rho_{jam}+1) }\right)\right.\nonumber\\ 
&\;\left.-\frac{T_d}{T}\log\left(1+M_e\rho_k+\frac{M_eM\rho_k\rho_{jam}}{\rho_rT_r+\rho_{jam}+1} \right)\right]^+\label{rate_11}
\end{align}
due to the fact that $L = T_r$.  Suppose $\gamma \leq 1$.  We bound $R_k$ as follows 
\begin{align}
&R_k\geq \frac{T_d}{T} \left[\log\left(1+\frac{M\rho_k \rho_r M^{\gamma}}{(\rho_f+\rho_{jam} +1)(\rho_rM^{\gamma}+\rho_{jam}+1) }\right)\right.\nonumber\\ 
&\quad\left.-\log\left(1+M_e\rho_k+\frac{M_eM\rho_k\rho_{jam}}{\rho_rM^{\gamma}+\rho_{jam}+1} \right)\right]^+\label{rate_12}\\
\geq &\frac{T_d}{T} \left[\log M+\log\left(\frac{\rho_k \rho_r }{(\rho_f+\rho_{jam} +1)(\rho_r+\rho_{jam}+1) }\right)\right.\nonumber\\ 
&\left.-(1-\gamma)\log M-\log\left(1+M_e\rho_k+\frac{M_e\rho_k\rho_{jam}}{\rho_r} \right)\right]^+\nonumber\\
&=\frac{T_d}{T} \left[\gamma\log M-\log\left(\left(1+M_e\rho_k+\frac{M_e\rho_k\rho_{jam}}{\rho_r}\right)\right.\right.\nonumber\\ 
&\quad\quad\quad\qquad\left.\left. \times(\rho_f+\rho_{jam} +1)\frac{\rho_r+\rho_{jam}+1}{\rho_k \rho_r}\right)\right]^+\label{rate_13}
\end{align}
where \eqref{rate_12} follows from the fact that $T_r \geq M^{\gamma}$. Notice that the second logarithm term in~\eqref{rate_13} does not depend on $M$.  Hence, we observe that $\frac{R_k}{\log M}\geq \frac{T_d}{T}\gamma-\epsilon$ if $M\geq G(\epsilon)$. In a similar way, for $\gamma >1$, we can show that  $\frac{R_k}{\log M}\geq \frac{T_d}{T}-\epsilon$ if $M\geq G(\epsilon)$.
\section{}
\subsection{Proof of Theorem~\ref{no_wyner2}}\label{proof_of_no_wyner2}
We set the size of the pilot signal set $L$ to $T_r$. Let $J$ be any integer in set $\{1,\dots,T_r\}$. The BS uses the same strategy explained in the proof of Theorem~\ref{encrypting} in order to estimate the gains of channels connecting the BS to users.

The BS picks arbitrary message rates $R_k>0$, $k=1,\ldots, K$. The equivocation rate for a code $\left(2^{BTR_1},\dots,2^{BTR_K}, BT_d\right)$ utilizing deterministic encoding mapping functions, $f_k$, $k=1,\ldots,K$ and $\delta$-conjugate beamforming is as follows:
\begin{align} 
&\frac{1}{BT} H\left(W_k| Z^{BT_d}, H^B,\hat{H}^B,H^B_e \right)\nonumber\\  
&\geq R_k-\frac{T_d}{T}\log\left(1+\frac{M_e\rho_k}{M^{\delta}}+\frac{M^{1-\delta}M_e\rho_k\rho_{jam}a}{L\rho_r} \right)\label{no_wyner2_1}\\
&=R_k-\frac{T_d}{T}\log\left(1+\frac{M_e\rho_k}{M^{\delta}}+\frac{M^{1-\delta}M_e\rho_k\rho_{jam}}{\rho_r T_r+T_r+1} \right)\label{no_wyner2_2}\\
&\geq R_k-\frac{T_d}{T}\log\left(1+\frac{M_e\rho_k}{M^{\delta}}+M^{1-\delta-\gamma}\frac{M_e\rho_k\rho_{jam}}{\rho_r } \right)\label{no_wyner2_3}
\end{align}
for all $k\in \{1,\dots,K\}$, where $a$ is defined in~\eqref{rate1111} and $\hat H_k(i)$ for any $k\in \{1,\dots,K\}$ and $i\in\{1,\ldots,B\}$ is given in~\eqref{chest}\color{black}. In the above derivation, \eqref{no_wyner2_1} follows from a derivation that is similar to \eqref{training11}-\eqref{independence11} in Appendix~\ref{no_wyner_proof}, \eqref{no_wyner2_2} follows from the fact the cardinality of pilot signal set $L$ is chosen as $T_r$ and \eqref{no_wyner2_3} follows from the fact that $T_r\geq  M^{\gamma}$.

As $\delta+\gamma >1$, the RHS of \eqref{no_wyner2_3} goes to zero as $M\to\infty$. For any $\epsilon>0$, $M\geq S_1(\epsilon)$ implies that right hand side of \eqref{no_wyner2_3} is smaller than $\epsilon$, completing the proof.

\subsection{Proof of Corollary~\ref{revisit1}}\label{proof_of_revisit1}
We set the size of the pilot signal set $L$ to $T_r$. Let $J$ be any integer in set $\{1,\dots,T_r\}$. The BS uses the same strategy explained in the proof of Theorem~\ref{encrypting} in order to estimate the gains of channels connecting the BS to users. 

Pick  $\delta$ and $\gamma$ such that  $0<\delta<1$ and $\gamma+\delta>1$.  Pick any arbitrary $\epsilon>0$ and arbitrary rate tuple $R=[R_1, \dots,R_K]$. Choose $M$ such that $M\geq \max(V_1(R),S_1(\epsilon))$. Note that inequality $M\geq V_1(R)$ implies that $R_k\leq \frac{T_d}{T} \log\left(1+\frac{M^{1-\delta}\rho_ka}{M^{-\delta}\rho_f+\rho_{jam} +1}\right)$ for all $k\in\{1,\ldots, K\}$, where $a$ is defined in~\eqref{rate1111}. As in the proof of Corollary~\ref{revisit}, we can show that there exists $B(\epsilon)>0$  and a sequence of  codes $\left(2^{BT R_1},\dots, 2^{BT R_K}, BT_d\right)$    that satisfy the decodability constraint in~\eqref{cond1} for $B\geq B(\epsilon)$, when $\delta$-beamforming, combined with deterministic mapping is used. In addition, since $M\geq S_1(\epsilon)$ and $T_r\geq M^\gamma$, following Theorem~\ref{no_wyner2}, the sequence of codes mentioned above satisfy the constraint in~\eqref{l3}, completing the proof.


\end{document}